\documentclass{aa}
\usepackage{graphicx}
\usepackage{txfonts}
\usepackage{natbib}
\usepackage{url} 
\begin{document}

%USER DEFINED SHORTCUTS
\newcommand{\zabs}{z_{\rm abs}}
\newcommand{\dla}{damped Lyman-$\alpha$}
\newcommand{\DLA}{Damped Lyman-$\alpha$}
\newcommand{\CI}{C\,{\sc i}}
\newcommand{\CII}{C\,{\sc ii}}
\newcommand{\CIV}{C\,{\sc iv}}
\newcommand{\SiIV}{Si\,{\sc iv}}
\newcommand{\HI}{H\,{\sc i}}  
\newcommand{\PII}{P\,{\sc ii}}
\newcommand{\SII}{S\,{\sc ii}}
\newcommand{\SiII}{Si\,{\sc ii}}
\newcommand{\FeII}{Fe\,{\sc ii}}
\newcommand{\ZnII}{Zn\,{\sc ii}}
\newcommand{\NiII}{Ni\,{\sc ii}}
\newcommand{\CrII}{Cr\,{\sc ii}}
\newcommand{\TiII}{Ti\,{\sc ii}}
\newcommand{\MnII}{Mn\,{\sc ii}}
\newcommand{\MgI}{Mg\,{\sc i}}
\newcommand{\MgII}{Mg\,{\sc ii}}
\newcommand{\PbII}{Pb\,{\sc ii}}
\newcommand{\CuII}{Cu\,{\sc ii}}
\newcommand{\lya}{Ly-$\alpha$}
\newcommand{\lyb}{Ly-$\beta$}
\newcommand{\lyg}{Ly-$\gamma$}
\newcommand{\ArI}{Ar\,{\sc i}}
\newcommand{\NI}{N\,{\sc i}}
\newcommand{\OI}{O\,{\sc i}}
\newcommand{\OVI}{O\,{\sc vi}}
\newcommand{\NV}{N\,{\sc v}}

%TITLE%%%%%%%%%%%%%%%%%%%%%%%%%%%%%%%%%%%%%%%%%%%%%%%%%%%%%%%%
\title{Excitation mechanisms in newly discovered H$_2$-bearing Damped
  Lyman-$\alpha$ clouds: systems with low molecular fractions
\thanks{Based on observations carried out at the European Southern Observatory (ESO)
under progs. ID 67.A-0022, 69.A-0204, 072.A-0346, 072.A-0442, 073.A-0071 and 074.A-0201 with the
UVES spectrograph installed at the Very Large Telescope (VLT) Kueyen
UT2 on Cerro Paranal, Chile}}
%\subtitle{Analysis of systems with low molecular fractions}
\titlerunning{%Two new H$_2$-detections in DLAs
Excitation mechanisms in newly discovered H$_2$-bearing DLAs: systems with low
molecular fractions. %amped Lyman-$\alpha$ systems
}
%%%%%%%%%%%%%%%%%%%%%%%%%%%%%%%%%%%%%%%%%%%%%%%%%%%%%%%%%%%%%%

%AUTHORS AND INSTITUTE%%%%%%%%%%%%%%%%%%%%%%%%%%%%%%%%%%%%%%%%%%%%%%%%%%%%%%%%%%%%%
\author{P. Noterdaeme\inst{1} \and C. Ledoux\inst{1} \and
  P. Petitjean\inst{2,3} \and F. Le Petit\inst{4} \and
  R. Srianand\inst{5} \and A. Smette\inst{1}}

\institute{European Southern Observatory, Alonso de C\'ordova 3107, Casilla 19001, Vitacura,
Santiago, Chile,
\email{pnoterda@eso.org} %, cledoux@eso.org, asmette@eso.org}
\and
Institut d'Astrophysique de Paris, CNRS - Universit\'e Pierre et Marie
Curie, 98bis Boulevard Arago, 75014, Paris, France
%\email{petitjean@iap.fr}
\and
LERMA, Observatoire de Paris, 61 Avenue de l'Observatoire, 75014, Paris, France
\and
LUTH, Observatoire de Paris, 61 Avenue de l'Observatoire, 75014, Paris, France
%\email{franck.lepetit@obspm.fr}
\and
IUCAA, Post Bag 4, Ganesh Khind, Pune 411\,007, India
%\email{anand@iucaa.ernet.in}
}

\date{Received ; accepted}
%%%%%%%%%%%%%%%%%%%%%%%%%%%%%%%%%%%%%%%%%%%%%%%%%%%%%%%%%%%%%%%%%%%%%%%%%%%%%%%%%%%

%ABSTRACT%%%%%%%%%%%%%%%%%%%%%%%%%%%%%%%%%%%%%%%%%%%%%%%%%%%%%%%%%%%%%%%%%%%%%%
\abstract
{}
%AIMS
{%We study the H$_2$ molecular fraction in the ISM of high-redshift DLA galaxies and
%derive the physical conditions in the gas.
We probe the physical conditions in high-redshift damped Lyman-$\alpha$ systems (DLAs) using the observed molecular fraction and the
rotational excitation of molecular hydrogen.
%We present two new detections of H$_2$ in damped Lyman-$\alpha$ (DLA) systems at
%  high redshift.
}
%METHOD
{We search for Lyman- and Werner-band absorption lines of molecular hydrogen in the
  VLT/UVES spectra of background QSOs at the redshift of known DLAs.}
%RESULTS
{We report two new detections of molecular hydrogen in the systems at 
$z_{\rm abs}$~=~2.402 and 1.989 toward, respectively, HE\,0027$-$1836 and HE\,2318$-$1107, discovered
in the course of the Hamburg-ESO DLA survey. We also present a
detailed analysis of our recent H$_2$ detection toward Q\,2343$+$125. 
  All three systems have low molecular fractions, $\log f \leq -4$, with
  $f=2N($H$_2)/(2N($H$_2)+N($\HI$))$. Only one such H$_2$ system was known previously.
Two of them  (toward  Q\,2343$+$125 and HE\,2318$-$1107) have high-metallicities,
  [X/H$]>-1$, whereas the DLA toward  HE\,0027$-$1836 is the system
  with the lowest metallicity ([Zn/H$]=-1.63$) among known H$_2$-bearing DLAs.
The depletion patterns for Si, S, Ti, Cr, Mn, Fe and Ni in the three 
systems are found to be very similar to what is observed in diffuse gas of the Galactic halo.
  Molecular hydrogen absorption from rotational levels up to J~=~5 is observed in a single
well-defined component toward HE\,0027$-$1836. We show that 
the width (Doppler parameter) of the H$_2$ lines increases with
increasing J and that the kinetic energy derived from the Doppler parameter is linearly dependent 
on the relative energy of the rotational levels.
There is however no velocity shift between lines from different rotational levels.
The excitation temperature is found to be 90~K for J~=~0 to J~=~2 and $\sim$500~K for higher J levels.
%very similarly to what is observed in some lines of sight of
%the galactic ISM. 
%Added Anand---:
Single isothermal PDR  models fail to reproduce the observed
rotational excitations. A two-component model is needed: one component
of low density ($\sim$50~cm$^{-3}$) with weak illumination ($\chi$~=~1)
to explain the J~$\le$~2 rotational levels and another of high density
($\sim$500~cm$^{-3}$) with strong illumination ($\chi$~=~30) for
J~$\ge$~3 levels.
% contributing to the absorption profile. 
However, the juxtaposition of these two PDR components may be ad-hoc and
the multicomponent structure could result either from turbulent dissipation or C-shocks.
%commented Anand---:
%Models support the idea that two gaseous components
%are present, one with small density ($n_{\rm H}$~$\sim$~10-50~cm$^{-3}$) 
%embedded in a UV radiation field with intensity similar to the Galactic one
%($\chi$~=~1), to explain the J~=~0-2 rotational level populations;
%and a second component with high density and intense UV field 
%($n_{\rm H}$~$\sim$~500~cm$^{-3}$ and $\chi$~=~30) to explain the
%J~$\ge 3$ populations. However, 
%turbulent dissipation is as likely a solution to explain
%the excitation of the molecular gas.
%----
}
{}
% for the first time at high-redshift a broadening of 
%the H$_2$ lines increasing with the rotational energy of the
%  molecules. 
%The ambient flux derived from the C~{\sc ii}$^*$ column density is twenty
%times larger than in our Galaxy and is consistent with the value derived
%from the observed excitation of H$_2$. 
%{These new detections increase to fifteen the number of known H$_2$-bearing 
%DLAs. The physical conditions found in these three DLAs with low molecular
%content strongly supports the idea that part of the DLA population
%is located in the halo of the associated galaxy probably expelled
%from the disk by strong galactic disks.
%
%and showed
%two mechanisms may be responsible for the J-dependent broadening in
%the system toward HE\,0027$-$1836. This can be explained by
%H$_2$-formation on dust, or by collisional excitation due to
%turbulences or C-shocks.
%


%%%%%%%%%%%%%%%%%%%%%%%%%%%%%%%%%%%%%%%%%%%%%%%%%%%%%%%%%%%%%%%%%%%%%%%%%%%%%%%%
\maketitle

\section{Introduction}
High-redshift \dla\ systems (DLAs) are the absorbers of highest
  \HI\footnote{To avoid confusion, we use the convenient
  spectroscopic notation (e.g. \SiII\ for Si$^+$) for atoms and ions 
  following the convention widely used in this field.} column density seen in QSO spectra, with $N($\HI$)\ge
2\times 10^{20}$ cm$^{-2}$. These absorbers are a major reservoir of
neutral hydrogen at $2\le z \le3$ \citep[e.g.][]{Prochaska05} and have long been identified as the
precursors of present day galaxies \citep[see][for a recent review]{Wolfe05}.
Our understanding of DLAs is primarily based on 
absorption-line studies, involving the detection of low-ionization
metal transitions and, in a few cases, molecular hydrogen
\citep[e.g.][]{Ledoux03, Srianand05}.
Molecular hydrogen is conspicuous in our Galaxy: lines of sight with $\log N$(\HI)~$\ge$~21
usually have $\log N$(H$_2$)~$\ge$~19 \citep{Savage77, Rachford02} whilst
high-redshift ($\zabs >1.8$) H$_2$-bearing DLAs are rare
with only ten detections reported up to now, 
with redshifts up to $\zabs=4.2$ \citep{Ledoux06b}. 
\citet{Petitjean06} noted a correlation between the
presence of molecular hydrogen and the metallicity ([X/H$]=\log
N($X$)/N($H$) -\log$~X/H$_\odot$, with
X~=~Zn, S or Si) in the gas that may
help select the best candidates for future surveys.
% using a threshold limit on the molecular
%fraction of $\log f \geq -4$ (with $f=2N($H$_2)/(2N($H$_2)+N($\HI$))$).
Indeed, 40\% of the [X/H$]>-1.3$ DLAs have $\log f > -4$,
with $f=2N($H$_2)/(2N($H$_2)+N($\HI$))$. On the other hand, only
15\% of the overall sample (with $-2.5<[$X/H$]<-0.3$) have $\log f > -4$. 
Furthermore, up to now, there is no H$_2$ detection in [X/H$]<-1.5$
DLAs, down to a detection limit of typically $N($H$_2)=2\times10^{14}$~cm$^{-2}$.
%
%\par\noindent
%
\par The distribution of molecular fractions is observed to be bimodal: 
all the upper limits are lower than $\log f = -4.5$,
whereas all detections above this limit have $\log f > -3$.
Low molecular fractions (with $\log f \le -4$) are basically unexplored
\citep[see however][]{Levshakov02}. 
Up to now, only two low-$f$ systems have been reported, in the DLA at
$\zabs=3.025$ toward Q\,0347$-$383 \citep[$\log f = -5.9$;][]{Levshakov02,Ledoux03} and 
in the DLA at
$\zabs=2.431$ toward Q\,2343$+$125 \citep[$\log f = -6.4$;][]{Petitjean06}.
%
%\par\noindent
%
%Both the excitation processes and the formation of H$_2$ are still
%poorly understood, despite extensive observational, theoretical and experimental
%investigations. 
\par Molecular hydrogen is primarily formed on the surface of dust grains in the interstellar
medium \citep[e.g.][]{Hollenbach71a}, but can also result from the
formation of negative hydrogen (H$^-$) if the gas is warm and dust-free
%\citep[e.g.][]{Cazaux02}.
{\citep[e.g.][]{Black87}}. 
{H$_2$ is also formed  in the gas phase by 
radiative association of H$^+$ with H to form H$_2^+$, 
followed by charge transfer with H. However, the reaction is slow in the
conditions prevailing in the DLA gas.}
\citet{Spitzer75} first showed that the observed relative 
populations of the
H$_2$ rotational levels cannot be described by a single excitation
temperature. Whilst the low-J levels can be thermalised,
the high-J levels are populated by direct formation on these levels
and UV pumping \citep[e.g.][]{Jura75} or by collisions if part
of the gas is heated {to temperatures above a few hundred K by,
for example,} turbulent dissipation and/or shocks \citep[e.g.][]{Joulain98, Cecchi-Pestellini05}. 
In DLAs, H$_2$ excitation is generally explained by predominant UV pumping \citep{Hirashita05}.
%
%\par\noindent
%
%At densities characteristic of diffuse clouds, the
%collision time scale fall between the low and high J radiative time. Low-J
%levels therefore represent roughly the kinetic temperature, while the
%high-J levels are not thermalized and reflects the efficiency of the
%excitation processes. High-J levels may be mainly populated by UV and
%formation on dust pumpings \citep[e.g.][]{Jura75} or by turbulences or
%shocks \citep[e.g.][]{Joulain98, Cecchi-Pestellini05}. In DLAs, H$_2$
%excitation can generally be explained by UV pumping
%\citep{Hirashita05}.
\par Here we present the analysis of H$_2$ absorptions in three DLAs, at $\zabs =2.431$, 
1.989, and 2.402 toward, respectively, Q\,2343$+$125, HE\,2318$-$1107, and HE\,0027$-$1836. The three systems have
low molecular fractions according to the above criterion, $\log f \le -4$.
The $\zabs=2.431$ system toward Q\,2343$+$125 has the lowest molecular fraction measured till now in
a DLA. We recently reported this detection
as part of a high-metallicity ([X/H$]> -1.3$) UVES DLA
sample \citep{Petitjean06}. A detailed analysis of the system is given here.
The latter two systems are new detections discovered during the Hamburg-ESO DLA
survey \citep[][Smette et al. 2007, in prep.]{Smette05}
and are reported for the first time.
%We focus on the 
H$_2$ is detected in rotational levels up to J~=~5 in the DLA toward 
HE\,0027$-$1836. This is the system with the lowest metallicity
amongst known H$_2$-bearing DLAs. % has ever been observed. 
%We are
%provided with very good data quality that allows us to clearly show
%for the first time at high-redshift, a broadening of the H$_2$-lines increasing with the
%rotational level J.
We present the observations in Sect.~2 and properties of the three systems toward
respectively Q\,2343$+$125, HE\,2318$-$1107 and HE\,0027$-$1836 in
Sects.~3 to 5. We discuss the H$_2$ excitation toward HE\,0027$-$1836 in Sect.~6
and comment on metallicity and depletion in Sect.~7. 
We conclude in Sect.~8.

%%%%%%%%%%%%%%%%%%
% Observations
%%%%%%%%%%%%%%%%%

\section{Observations}
The quasars Q\,2343$+$125, HE\,2318$-$1107, and HE\,0027$-$1836 were
observed in visitor mode with the Ultraviolet
and Visible Echelle Spectrograph \citep[UVES,][]{Dekker00}
mounted on the ESO Kueyen VLT-UT2 8.2\,m telescope
on Cerro Paranal in Chile. Q\,2343$+125$ was observed on October
29, 2003 under program ID.~072.A-0346. HE\,2318$-$1107 and HE\,0027$-$1836 were observed on September 16, 17, and 18, 2004, under program ID.~073.A-0071. HE\,0027$-$1836 was
re-observed on October 8 and 9, 2004, under program
ID.~074.A-0201. Observations were supplemented with UVES data from the
ESO archive, progs. 67.A-0022 and 69.A-0204 (P.I: D'Odorico) for
Q\,2343$+$125 and prog. 072.A-0442 (P.I: Lopez) for HE\,2318$-$1107
and HE\,0027$-$1836.
The data were reduced using the UVES pipeline \citep{Ballester00}
which is available as a context of the ESO MIDAS data reduction
system.
The main characteristics of the pipeline are to
perform a precise inter-order background subtraction, especially for master
flat-fields, and to allow for an optimal extraction of the object
signal rejecting cosmic rays and performing sky subtraction at the same time.
The pipeline products were checked step by step.
The wavelength scale of the reduced spectrum was then converted to vacuum-heliocentric 
values and the portions of the spectrum corresponding to different settings were each 
rebinned to a constant wavelength step. No further
rebinning was performed during the analysis of the whole spectrum.
Individual 1-D exposures were scaled, weighted and combined together. Standard
Voigt-profile fitting methods were used for the analysis to determine column
densities using the vacuum wavelengths and oscillator strengths from
\citet{Morton03} for metal ions, except for the oscillator strengths of
\FeII\,$\lambda\lambda\lambda\lambda$1063,1064,1112,1121 \citep{Howk00} and \NiII\,$\lambda\lambda$1317,1370 \citep{Jenkins06}, and the
wavelengths of \SII\,$\lambda\lambda$1250,1253 \citep{Morton91}. 
We used the oscillator strengths from the Meudon
group\footnote{\url{http://amrel.obspm.fr/molat}} based on
calculations described in  \citet{Abgrall94} for H$_2$.
%with updated wavelengths from either \citet{Philip04} or
%\citet{Ubachs04} (when not in \citet{Philip04}). 
We adopted the Solar abundances given by \citet{Morton03} based on the
  meteoritic data from \citet{Grevesse02}.
{For each system, the origin ($v = 0$~km\,s$^{-1}$) of the velocity scale is set to the redshift of the
  H$_2$ component.}

%\section{New H$_2$ detections}% in two high-metallicity DLAs, toward Q\,2343$+$125 and HE\,2318$-$1107}
%We present in this section the detection of molecular hydrogen at
%$\zabs=2.431$ toward Q\,2343$+$125, which we have annouced in
%\citet{Petitjean06}, at $\zabs=1.989$ toward HE\,2318$-$1107 and
%$\zabs=2.402$ toward HE\,0027$-$1836 which are reported here for the first time.

%%%%%%%%%%%%%%%%%%%%%%%%%
%%% Q2343
%%%%%%%%%%%%%%%%%%%%%%%%%
\section{The system at $\zabs=2.431$ toward Q\,2343$+$125}
\begin{figure}[!h]
 \begin{center}
 \includegraphics[clip=,width=0.89\hsize]{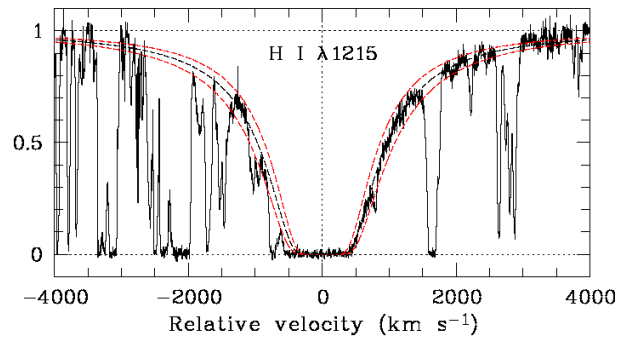}
 \caption{\label{Q2343_HI} Fit to the \dla\ line at $\zabs=2.431$ toward
 Q\,2343$+$125. The derived neutral hydrogen column density is $\log
 N$(\HI)~$=20.40\pm0.07$. {The dashed curves correspond to the
 best-fitted Voigt profile (black) and the associated errors (red).}}
 \end{center}
\end{figure}

%\subsection{The system at $\zabs=2.431$ toward Q\,2343$+$125}
This DLA has first been studied by \citet{Sargent88}, then by
\citet{Lu96}, \citet{Dodorico02} and \citet{Dessauges-Zavadsky04} with high
spectral resolution ($\sim 6.5$~km\,s$^{-1}$). We recently reported the detection of molecular
hydrogen in this system \citep{Petitjean06}. From Voigt-profile fitting to the \HI\ \lya, $\beta$ and $\gamma$
lines, we find that the \dla\ component is centered at $\zabs=2.431$ and the
column density is $\log N$(\HI)~$=20.40\pm0.07$, consistent with
previous measurement by \citet{Dodorico02}: $\log
N$(\HI)~$=20.35\pm0.05$. The fit to the damped \lya\ line is shown in Fig.~\ref{Q2343_HI}.
\subsection{Molecular hydrogen and neutral carbon content}
\begin{figure*}[!ht]
 \begin{center}
 \includegraphics[angle=90,width=\hsize]{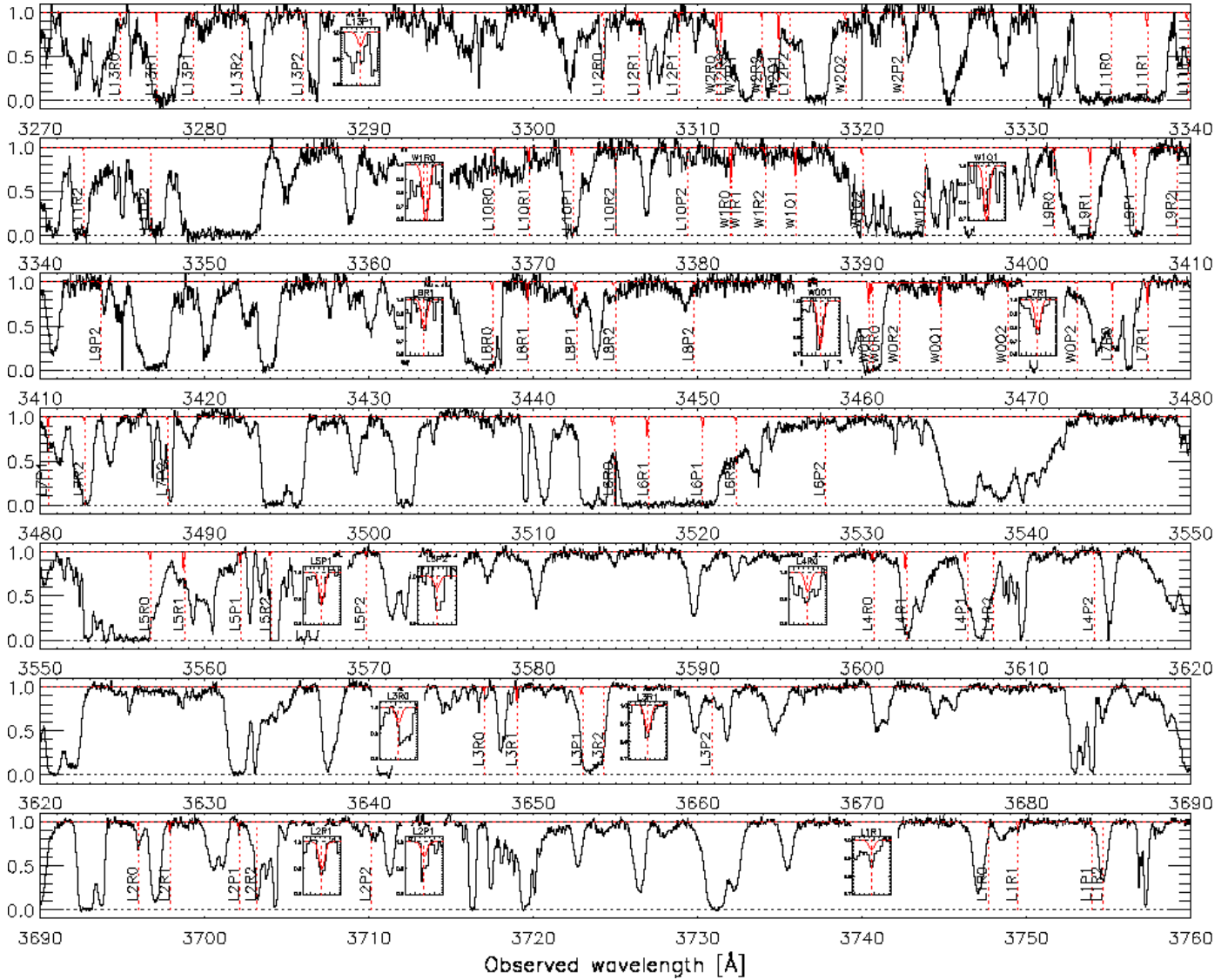}
 \caption{\label{q2343_H2} Portion of the normalized spectrum of
Q\,2343$+$125. The resolving power is $R\simeq 45\,000$.  The
best fit to the H$_2$ lines is superimposed on the
spectrum. Transition names are {written AxBz, with A~=~L or W for
  the Lyman or Werner band systems; x is the quantum number
of the vibrational upper state, while B is the branch designation
(P, Q or R for $\Delta$J~=~+1, 0, $-$1 respectively) and z is the rotational quantum number 
of the lower state.} 
%with  standing for H$_2$\,Lx--0Yz. 
The fit parameters are
given in Table~\ref{q2343H2CI}. Zooms into individual lines are
shown in small boxes that are shifted horizontally for
clarity. Each box is 0.5~${\rm \AA}$ wide.}
 \end{center}
\end{figure*}

\begin{figure}[!ht]
 \begin{center}
 \includegraphics[clip=,width=0.98\hsize]{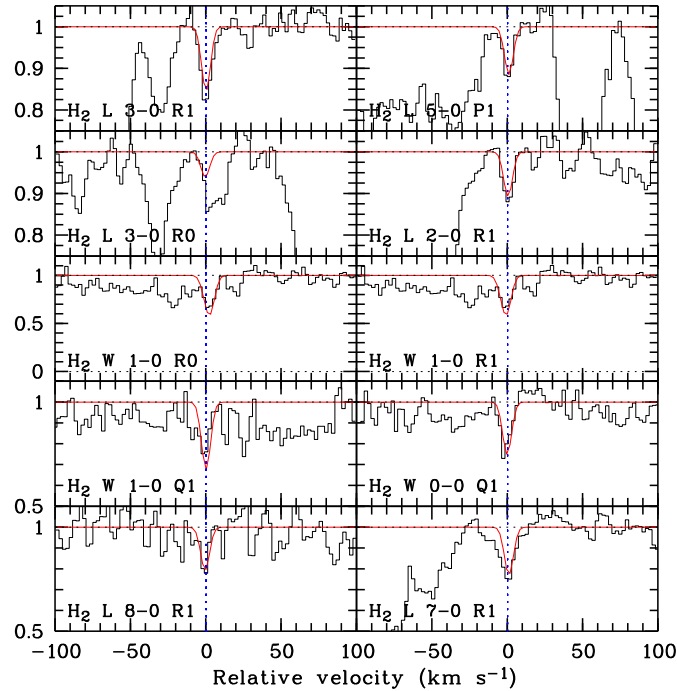}
 \caption{\label{q2343_H2_vel} Velocity plots of observed H$_2$
 absorption lines from the J~=~0 and J~=~1 rotational levels at
 $\zabs=2.43128$ toward Q\,2343$+$125. The model fit is
 over-plotted.}
 \end{center}
\end{figure}

%\subsection{H$_2$ and \CI\ contents}
The optically thin H$_2$ absorption lines at $\zabs =
2.43128$ are very weak, i.e. close
to but above the 3\,$\sigma$ detection limit (see Fig.~\ref{q2343_H2}). A very careful
normalization of the spectrum has been performed, adjusting the
continuum locally while fitting the line profiles. 
We detect the first two rotational levels in this system (see Fig.~\ref{q2343_H2_vel}). %{\bf PPJ: Remettre la figure ici - PN:done}.
%
%(See Fig.~\ref{q2343_H2}, for a velocity plot of H$_2$
%absorptions, see \citet{Petitjean06}). 
\noindent The total H$_2$ column density,
integrated over the J~=~0 and J~=~1 rotational levels, is measured to
be $\log N$(H$_2)=13.69\pm0.09$ ($12.97\pm0.04$ and $13.60\pm0.10$
for J~=~0 and J~=~1 respectively). This leads to an extremely small
molecular fraction $\log f=\log
2N$(H$_2$)/($2N$(H$_2$)+$N$(\HI$))=-6.41\pm0.16$.
We also derive a 3\,$\sigma$ upper limit on the detection of the J~=~2 level,
$\log N$(H$_2$, J=2)~$<13.1$, see Table~\ref{q2343H2CI}.
Neutral carbon is not detected resulting in an upper limit at 3\,$\sigma$:  $\log N$(\CI)~$<12.1$. This leads to a ratio
$\log N$(\CI)$/N$(H$_2)< -1.6$. 
\begin{table}[!ht]
\caption{\label{q2343H2CI} 
H$_2$ and \CI\ column densities in the $\zabs=2.431$ DLA toward Q\,2343$+$125.
%Results of Voigt-profile fitting for different H$_2$
%  rotational levels and upper-limits on the column densities of \CI\
%  fine-structure levels toward Q\,2343$+$125.
}
%\scriptsize
\begin{center}
\begin{tabular}{l l  l  l  c  c }
\hline
\hline

\#$^a$ & $\zabs$     &   level   & $\log N($H$_2$, J)$^b$  & $b^c$
   & $T_{\rm 0-J}^{~~~~d}$  \\
   &             &           &               & [km\,s$^{-1}$]  & [K]          \\
\hline
3  & 2.43128(3)  &   J = 0   &$12.97\pm0.04$ & 1.3-2.6    &  ---          \\
   &             &   J = 1   &$13.60\pm0.10$ &       ''       &  $229^{+173}_{-69}$          \\
   &             &   J = 2   &$<$~13.10$^e$      &       ''       &  $< 420$             \\

\hline
   & & & & & \\
\hline
\hline
\# &  $\zabs$    &  $\log N$(\CI) & $\log N$(\CI$^*$) & $b$ &$\Delta v_{\rm \ion{C}{i}/ H_2}$ \\
   &             &           &               & [km\,s$^{-1}$]  &  [km\,s$^{-1}$]         \\
\hline
   & 2.431       &  $<12.10^e$       & $<12.40^e$           & --- &          ---                    \\
\hline
\end{tabular}

\footnotesize
\end{center}
$^a$ The numbering refers to that of the metal components (see
Table~\ref{q2343metalst}).\\
$^b$ Errors represent the allowed range for $\log N$(H$_2$,J), not the rms
error from the fit.\\
$^c$ Depending on the inital guess, the fit converges to two values of $b$, we therefore give the
corresponding allowed range.\\
$^d$ The excitation temperatures between rotational levels 0 and J are
calculated using the allowed range for $N$(H$_2$,J).\\ 
$^e$ 3\,$\sigma$ upper limit.
\normalsize

\end{table}

%CI (upper a 3sig a 12.1)
%CI* < 12.4
%CI** < 12.3

%
\subsection{Metal content \label{parq2343m}}
\begin{figure}[!ht]
\begin{center}
\begin{tabular}{c}
\includegraphics[width=0.92\hsize,clip=]{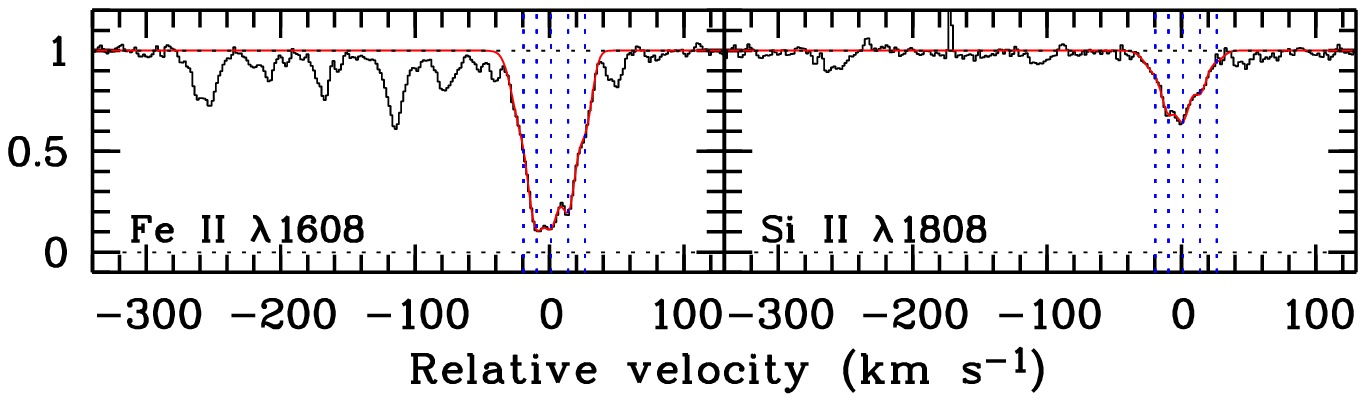}\\
\includegraphics[width=0.92\hsize,clip=]{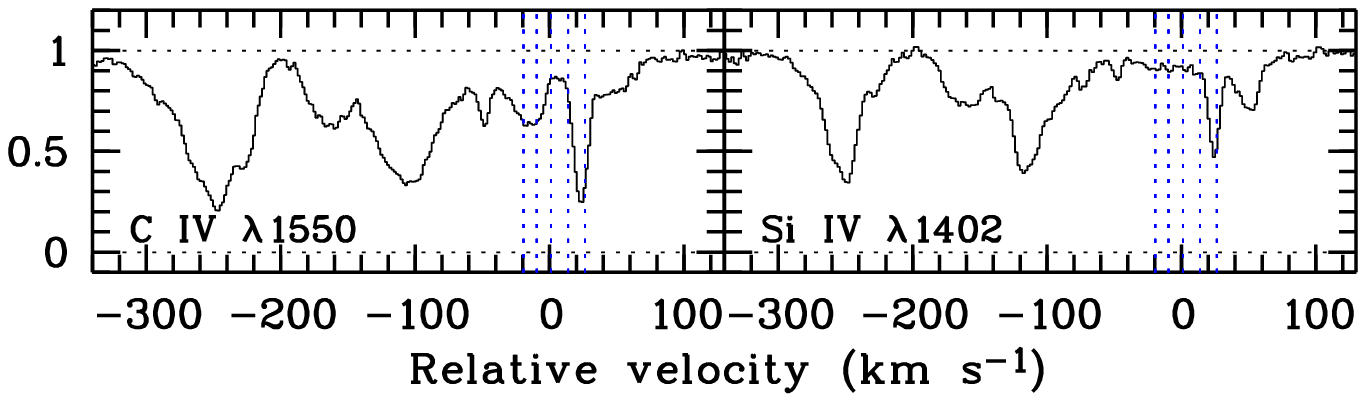}\\
\end{tabular}
\caption{\label{q2343CIVSiIV} Overall \SiIV\ and \CIV\ absorption
profiles in the $\zabs=2.431$ DLA toward Q\,2343$+$125. The vertical lines
  mark the redshift of the main low-ionization metal components (see Table~\ref{q2343metalst}). Note
  that the \CIV\ and \SiIV\ absorptions from the H$_2$-bearing
  component (at $v$~=~0~km\,s$^{-1}$) are absent (or weak). %corresponds to very
%  weak \CIV\ and \SiIV\ absorptions.
}
\end{center}
\end{figure}
\begin{figure}[!ht]
 \begin{center}
 \includegraphics[width=0.93\hsize,clip=]{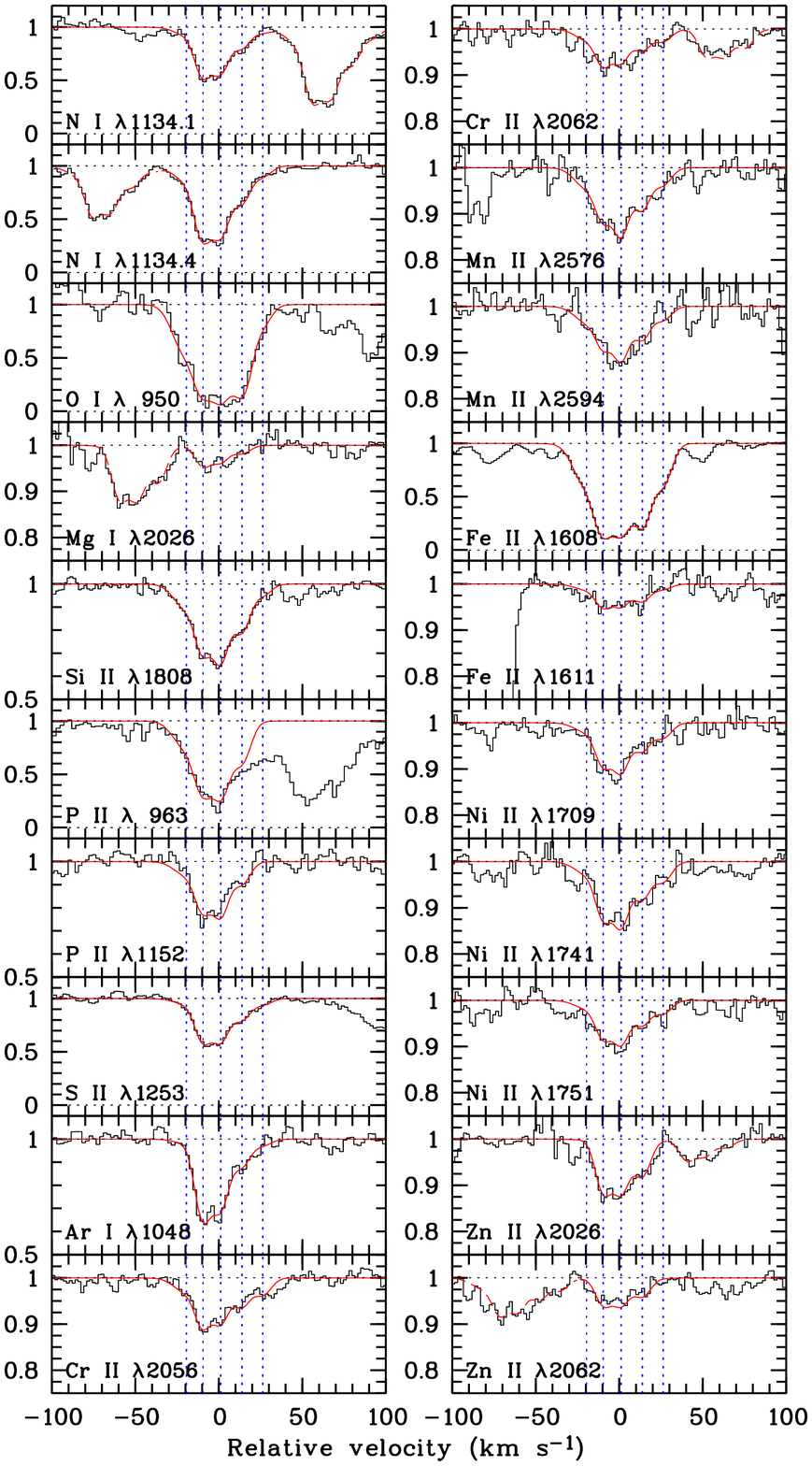}
\caption{\label{q2343_metals} Absorption profiles of low-ionization
  and neutral species in the \dla\ 
system at $\zabs=2.431$ toward Q\,2343$+$125.
%  Absorptions from \ZnII\,$\lambda$2026 and \MgI\,$\lambda$2026
%  appear on both panels, as well as \ZnII\,$\lambda$2062 and \CrII\,$\lambda2062$.
%Zero velocity is fixed at the position of the H$_2$ absorption at
%$z_{\rm abs}$~=~2.43128. 
The dotted profiles in some panels mark
absorption from another element.} 
 \end{center}
\end{figure}
\noindent The low-ionization metal profile is spread over more than 250~km\,s$^{-1}$
(Fig.~\ref{q2343CIVSiIV}) but shows a strong low-ionization clump
on its red side. We assume that most of the metals are in this 
clump as it is associated with by far the strongest \OI\ absorption
feature. 
%We choose to focus only on this strong absorption, that corresponds
%to the \HI- and also far the strongest \OI- absorptions. The ionization
%potential of \OI\ (13.62 eV) is very close to that of \HI\ (13.60 eV)
%and should therefore be a good tracer of neutral hydrogen. 
This main clump is itself decomposed into five components, among which the third 
one from the blue, at $\zabs = 2.43129$, shows associated H$_2$ absorption. About
35\% of \SiII\ and \SII\ reside in this component.
The Voigt-profile fitting is shown on Fig.~\ref{q2343_metals} and the
corresponding parameters are given in Table~\ref{q2343metalst}.
%The dust-free diagnostic tool for $\alpha$-enhancement, [O/Zn$]$ \citep[][]{Molaro00} 
%is about [O/Zn$]\sim 0.12 \pm 0.11$, averaging the three components where \ZnII\ is detected. This implies low $\alpha$-enhancement.
Note that there is very little \CIV\ and \SiIV\ absorption at the velocity
where low-ionization metal absorptions are the strongest (see Fig.~\ref{q2343CIVSiIV}).
High-ionization species are seen mostly blueshifted up to 300~km\,s$^{-1}$ from 
the main low-ionization clump. It is apparent that the profile is much
smoother far away from this component, suggesting a hot gas
outflow. 
%P.N: Following not true...(thx to Andrew!)...was a confusion.
%Note that \OVI\ and \NV\ are probably detected in this system
%\citep{Fox07}, and the \OVI\ profile is smoother than the
%\CIV\ and \SiIV\ profiles.
%
%
\begin{table}[!Ht]
\caption{\label{q2343metalst}
Low-ionization metals column densities in the $\zabs=2.431$ DLA toward Q\,2343$+$125.
% Voigt-profile
%  fitting results for low-ionization lines from the $\zabs=2.431$ DLA
%  toward Q\,2343$+$125.
}
\scriptsize
\begin{center}
\begin{tabular}{r l l l c c}
\hline
\hline
\# & $\zabs$   & Ion (X)   & $\log N$(X)     &  $b$              & $\Delta v_{\rm X/H_2}$ \\
   &           &           &                 &  [km\,s$^{-1}$] & [km\,s$^{-1}$]\\
%   &           &           &                  &                  & $\Delta v_{\rm X/\ion{C}{i}}$ \\
\hline

1   & 2.43105(7) & \NI\      & $13.49\pm0.02$  & $9.3\pm1.0$ &  ---\\
    &            & \OI\      & $15.46\pm0.02^a$& ''    & \\
    &            & \MgI\     & $<$~11.55$^b$   & ''    & \\
    &            & \SiII\    & $14.31\pm0.08$  & ''    & \\
    &            & \PII\     & $12.16\pm0.09$  & ''    & \\
    &            & \SII\     & $13.41\pm0.13$  & ''    & \\
    &            & \ArI\     & $11.64\pm0.26$  & ''    & \\
    &            & \CrII\    & $12.00\pm0.09$  & ''    & \\
    &            & \MnII\    & $11.51\pm0.10$  & ''    & \\
    &            & \FeII\    & $13.57\pm0.08$  & ''    & \\
    &            & \NiII\    & $12.39\pm0.14$  & ''    & \\
    &            & \ZnII\    & $<$~11.00$^b$   & ''    & \\

2   & 2.43116(9) & \NI\      & $14.20\pm0.01$  & $5.1\pm0.4$ &  ---\\
    &            & \OI\      & $\ge$~15.72$^c$ & ''    & \\
    &            & \MgI\     & $11.94\pm0.06$  & ''    & \\
    &            & \SiII\    & $14.55\pm0.07$  & ''    & \\
    &            & \PII\     & $12.53\pm0.06$  & ''    & \\
    &            & \SII\     & $14.20\pm0.04$  & ''    & \\
    &            & \ArI\     & $12.82\pm0.02$  & ''    & \\
    &            & \CrII\    & $12.31\pm0.06$  & ''    & \\
    &            & \MnII\    & $11.67\pm0.08$  & ''    & \\
    &            & \FeII\    & $13.96\pm0.05$  & ''    & \\
    &            & \NiII\    & $12.84\pm0.06$  & ''    & \\
    &            & \ZnII\    & $11.71\pm0.04$  & ''    & \\

3   & 2.43129(3) & \NI\      & $14.19\pm0.01$  & $5.6\pm0.5$ & $+0.8$\\
    &            & \OI\      & $\ge$~15.92$^c$       & ''  &  \\	  
    &            & \MgI\     & $11.87\pm0.08$  & ''  &  \\
    &	         & \SiII\    & $14.70\pm0.04$  & ''  &  \\
    &	         & \PII\     & $12.64\pm0.04$  & ''  &  \\
    &	         & \SII\     & $14.24\pm0.04$  & ''  &  \\
    &            & \ArI\     & $12.78\pm0.02$  & ''  &  \\
    &	         & \CrII\    & $12.34\pm0.05$  & ''  &  \\
    &	         & \MnII\    & $11.89\pm0.04$  & ''  &  \\
    &            & \FeII\    & $14.00\pm0.04$  & ''  &  \\
    &	         & \NiII\    & $12.97\pm0.04$  & ''  &  \\
    &	         & \ZnII\    & $11.74\pm0.04$  & ''  &  \\

4   & 2.43143(8) & \NI\      & $13.78\pm0.01$  & $5.6\pm0.5$ & --- \\
    &            & \OI\      & $\ge$~15.81$^c$       & ''   & \\
    &            & \MgI\     & $11.57\pm0.13$  & ''   & \\
    &	         & \SiII\    & $14.45\pm0.04$  & ''   & \\
    &	         & \PII\     & $12.22\pm0.06$  & ''   & \\
    &	         & \SII\     & $13.88\pm0.04$  & ''   & \\
    &            & \ArI\     & $12.32\pm0.04$  & ''   & \\
    &	         & \CrII\    & $12.14\pm0.05$  & ''   & \\
    &	         & \MnII\    & $11.66\pm0.05$  & ''   & \\
    &            & \FeII\    & $13.90\pm0.03$  & ''   & \\
    &	         & \NiII\    & $12.71\pm0.05$  & ''   & \\
    &	         & \ZnII\    & $11.57\pm0.04$  & ''   & \\

5   & 2.43158(1) & \NI\      & $13.18\pm0.03$  &  $5.8\pm0.4$ & ---\\
    &            & \OI\      & $14.82\pm0.06^a$  &  '' &  \\
    &            & \MgI\     & $<$~11.55$^b$   & ''  &  \\
    &	         & \SiII\    & $13.80\pm0.09$  & ''  &  \\
    &	         & \PII\     & $<$~11.85$^b$   & ''  &  \\
    &	         & \SII\     & $13.33\pm0.08$  & ''  &  \\
    &            & \ArI\     & $11.69\pm0.17$  & ''  &  \\
    &	         & \CrII\    & $11.92\pm0.05$  & ''  &  \\
    &	         & \MnII\    & $11.28\pm0.09$  & ''  &  \\
    &            & \FeII\    & $13.40\pm0.04$  & ''  &  \\
    &	         & \NiII\    & $12.45\pm0.06$  & ''  &  \\
    &	         & \ZnII\    & $<$~11.00$^b$   & ''  &  \\

\hline
\end{tabular}
\end{center}
\footnotesize
%$^a$ km\,s$^{-1}$\\
$^a$ \OI\ has been fitted using fixed Doppler parameters previously
derived from the other ions. Errors on $N$(\OI) could therefore be
underestimated. \\
$^b$ Upper-limits are given for non-detections at the 3\,$\sigma$ confidence level.\\
$^c$ $N$(\OI) is considered as a lower-limit in the three saturated components because of uncertainties in the background subtraction.

\normalsize
\end{table}

%%%%%%%%%%%%%%%%%%%%%%%%%%%%%%%%%%%%%%
%% HE2318
%%%%%%%%%%%%%%%%%%%%%%%%%%%%%%%%%%%%%

\section{The system at $\zabs=1.989$ toward HE\,2318$-$1107}
\par
\begin{figure}[!ht]
 \begin{center}
 \includegraphics[clip=,width=0.9\hsize]{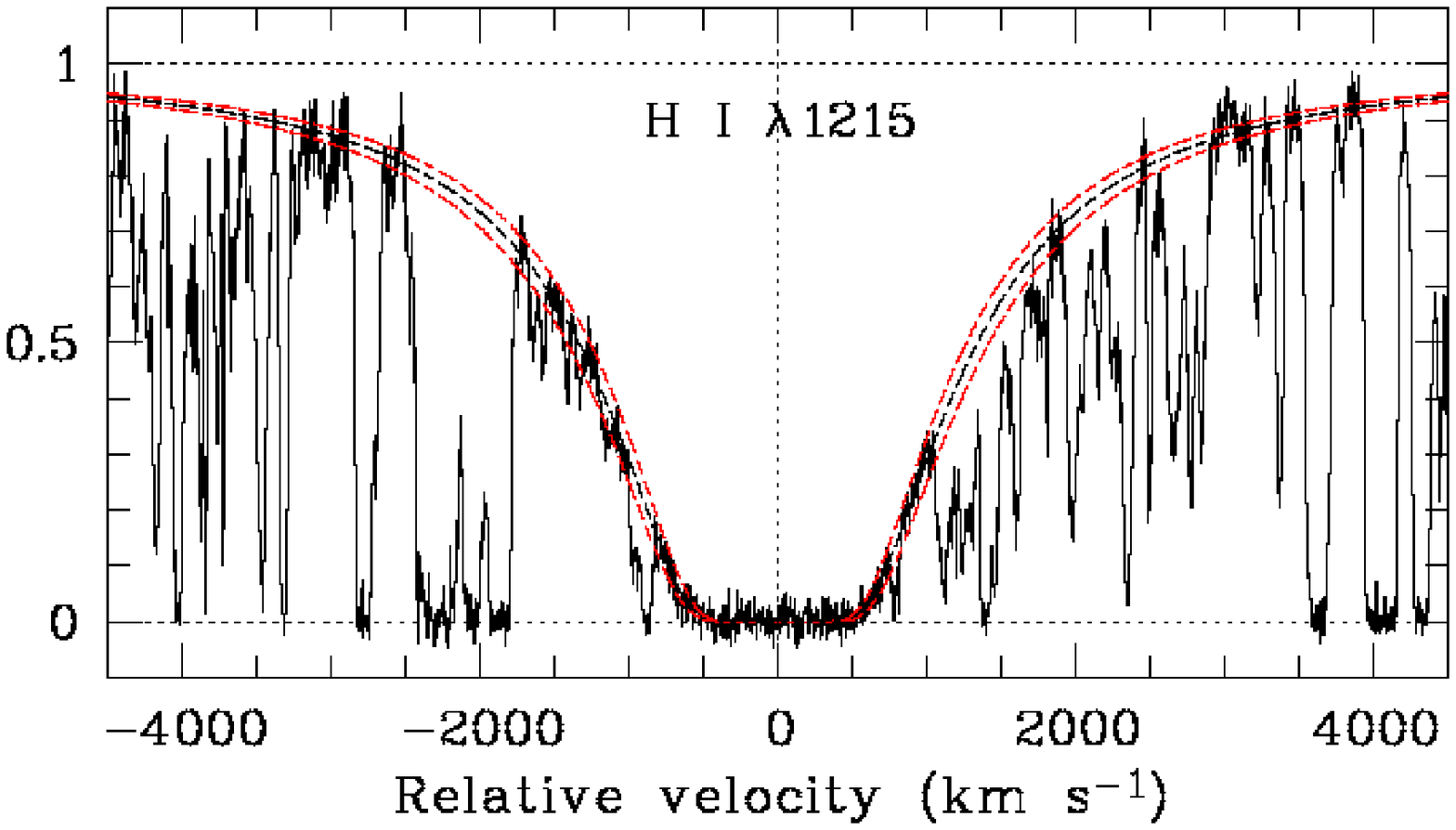}
 \caption{\label{HE2318_HI} Fit to the \dla\ line at $\zabs=1.989$ toward
 HE\,2318$-$1107. The derived neutral hydrogen column density is $\log
 N$(\HI)~$=20.68\pm0.05$. {The dashed curves correspond to the
 best-fitted Voigt profile (black) and the associated errors (red).}}
 \end{center}
\end{figure}

\noindent From Voigt-profile fitting to the \HI\ \lya\ and \lyb\ lines, we
derive a total neutral hydrogen column density of $\log
N$(\HI)~=~$20.68\pm0.05$. The fit is shown on Fig.~\ref{HE2318_HI}.
\subsection{Molecular content}
\begin{figure}[!ht]
 \begin{center}
 \includegraphics[clip=,width=\hsize]{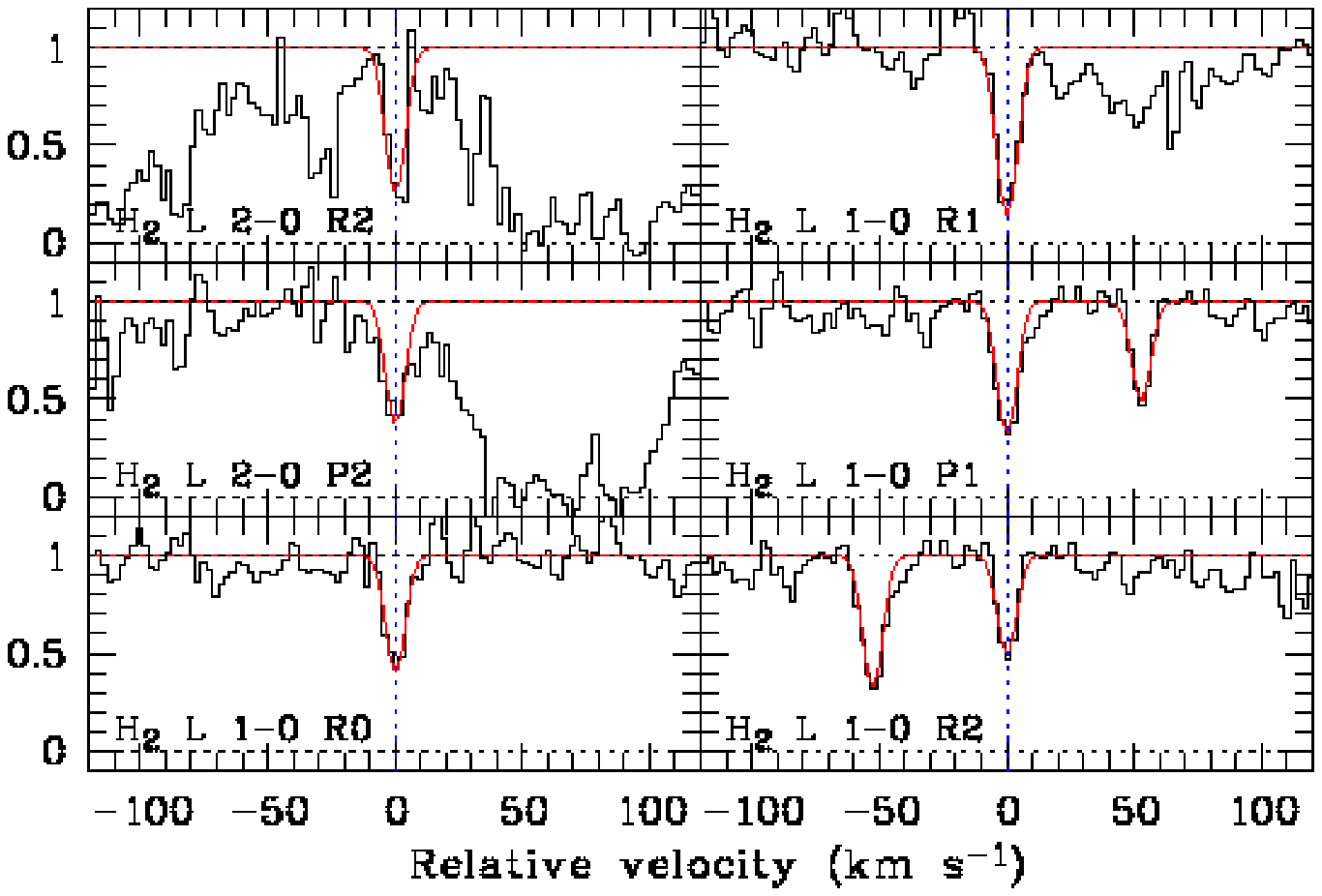}
 \caption{H$_2$ absorption profile at $\zabs=1.98888$ toward
 HE\,2318$-$1107. The model fit is over-plotted.\label{HE2318H2}}
 \end{center}
\end{figure}

Our spectrum of HE\,2318$-$1107 covers only a few H$_2$ lines, but
these are unambiguously detected in a single component at
$\zabs=1.98888$ with absorption lines from the rotational levels J~=~0
to J~=~2 (see Fig.~\ref{HE2318H2}). Some 
lines  from the J~=~3 level may be present but they are always blended or
in regions of low SNR. We therefore derived only an upper limit at 3\,$\sigma$: 
$\log N$(H$_2$, J=3)~$<14.7$. The results of the fit are given in Table~\ref{HE2318H2CI}.
The lines are optically thin and the quality of the spectrum is high enough to measure accurate
column densities in each of the detected levels. The normalization of
the R~$\sim$~54\,000 spectrum has
been done locally around each line. 
The total H$_2$ column density is
$\log N$(H$_2$)~=~$15.49\pm0.03$, leading to a low molecular
fraction $\log f = -4.89\pm0.08$. 
%We can estimate the kinetic
%temperature of the absorbing cloud, using the excitation temperature
%of rotational level J~=~1 \citep{Roy06},
Because J~=~1 is mainly populated by {reactive} collisions, the kinetic temperature should be close to the 
excitation temperature $T_{01} \simeq 188$~K \citep{Roy06}. 
%This is nevertheless uncertain, because of
%the J~=~1 level may not be thermalised, in which case it is a
%lower-limit, and because of the selective self-shielding of J~=~1,
%which makes $T_{01}$ an upper limit.
%
%
\addtolength{\tabcolsep}{-1pt}
\begin{table}[!ht]
\caption{\label{HE2318H2CI}
 H$_2$ and \CI\ column densities in the $\zabs=1.989$ DLA toward HE\,2318$-$1107.
%Voigt-profile fitting results for
%  different H$_2$ rotational levels and \CI\ ground-state fine-structure levels in the $\zabs=1.989$ \dla\ system toward
%  HE\,2318$-$1107.
}
 \begin{center}
  \begin{tabular}{p{0.2cm} p{1.3cm} l l c c}
\hline
\hline
\#$^a$ & $\zabs$  & level & $\log N$(H$_2$, J)$^b$ & $b$           &$T_{\rm 0-J}^c$ \\
   &          &       &                 &[km\,s$^{-1}$]& [K]       \\

\hline
5   & 1.98888(2)& J~=~0   & $14.72\pm0.03$  & $3.6\pm0.2$   & ---               \\
    &          & J~=~1   & $15.28\pm0.03$  & ''          & 188$^{+158}_{-59}$ \\
    &          & J~=~2   & $14.84\pm0.03$  & ''          & 384$^{+173}_{-91}$ \\
    &          & J~=~3   & $<$~14.7$^d$    & ''          & $<$~355            \\
\hline
   &          &       &                 &             &                    \\
\hline
\hline
\#$^a$ & $\zabs$    &$\log N$(\CI)  & $\log N$(\CI$^*$) &    $b$         & $\Delta v_{\ion{C}{i}/{\rm H_2}}$\\
   &            &               &                   &[km\,s$^{-1}$]& [km\,s$^{-1}$]     \\
\hline
5  & 1.98889(3) &$12.63\pm0.02$ & $12.30\pm0.04$    &  $2\pm1$   &  $+$1.1              \\
\hline
  \end{tabular}
 \end{center}
\footnotesize
$^a$ The numbering refers to that of the metal components (see Table \ref{HE2318_metalsT}).\\
$^b$ The errors on the H$_2$ column densities are the 1\,$\sigma$
uncertainty from fitting the Voigt-profiles.\\
$^c$ The excitation temperatures are calculated using the 3\,$\sigma$ range for H$_2$.\\
$^d$ 3\,$\sigma$ upper-limit.
\normalsize
\end{table}
\addtolength{\tabcolsep}{1pt}
\subsection{Neutral carbon}
\begin{figure}[!ht]
 \begin{center}
 \includegraphics[clip=,width=0.6\hsize]{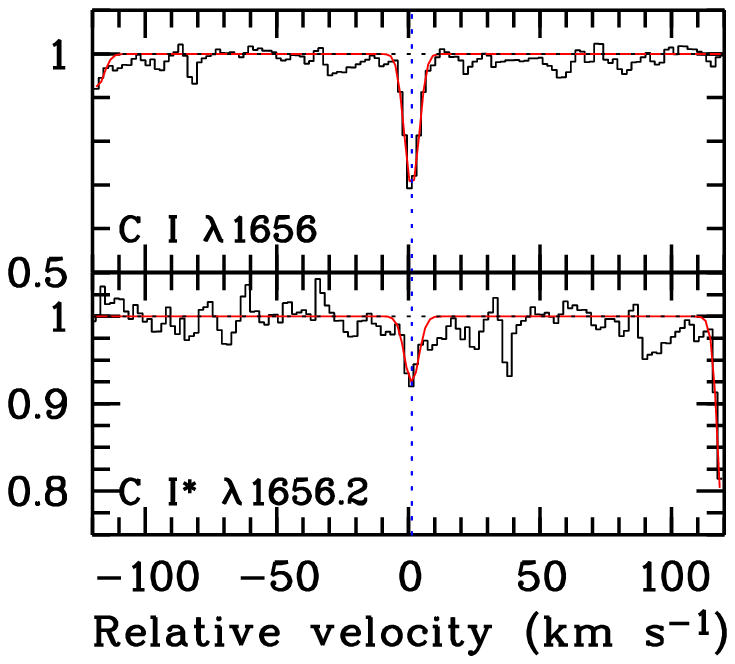}
 \caption{\CI\ ground-state fine-structure lines from the $\zabs=1.989$ DLA toward
 HE\,2318$-$1107. \label{HE2318CI}}
 \end{center}
\end{figure}

Neutral carbon is detected in two fine-structure levels of the ground state,
2s$^2$2p$^2$\,$^3$P$_0$ (\CI) and 2s$^2$2p$^2$\,$^3$P$_1$ (\CI$^*$), at the same redshift 
as the H$_2$ component (see Fig.~\ref{HE2318CI}). A velocity shift between H$_2$ and \CI\ of
only 1.1~km\,s$^{-1}$ is seen (see Table~\ref{HE2318H2CI}). This supports the idea that neutral carbon and molecular 
hydrogen absorptions originate from the same cloud.
%This is indeed not surprising, as the ionization potential of \CI\ (11.2\,eV) is very close 
%to the energy of the photons that excite H$_2$. 
Thanks to the \CI\ fine-structure level ratios, we can estimate the
particle density in the corresponding cloud. Using a simple {\sc popratio} model \citep{Silva01}, taking into account the cosmic background
radiation at redshift $z\sim2$ and using kinetic temperatures between $T=188$~K and 
$T=300$~K and a radiation field intensity 
between 1 and 10 times the Galactic value, leads to densities between 20
and 40~cm$^{-3}$. A higher UV flux would result in a lower density,
whereas a lower temperature would lead to a higher density, that would
be about twice higher than derived above for the mean Galactic temperature \citep[$T\sim77$~K;][]{Rachford02}.

\subsection{Metal content}
\begin{figure}[t]
 \begin{center}
 \includegraphics[clip=,width=\hsize]{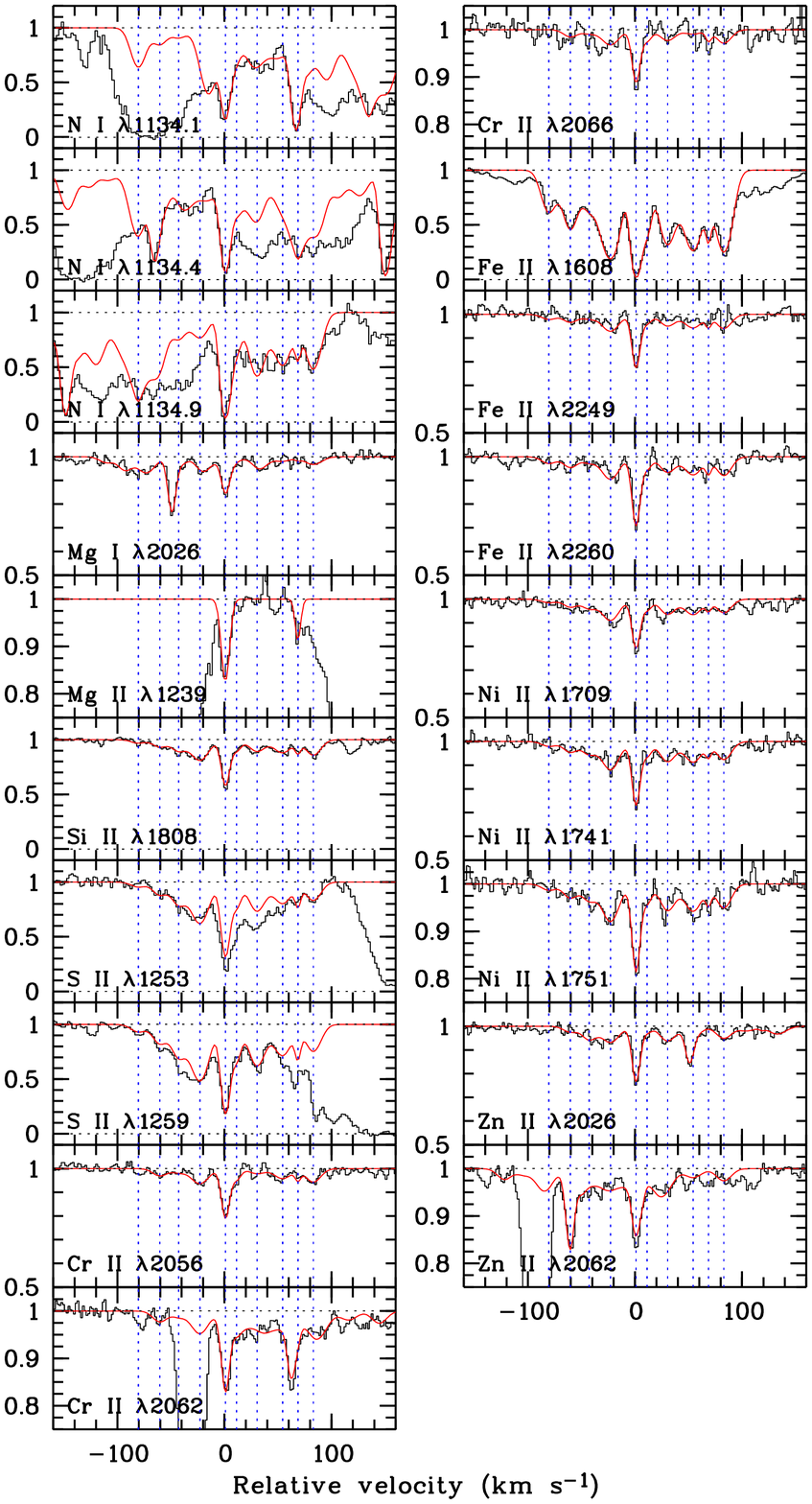}
 \caption{\label{HE2318metals} Absorption profiles of low-ionization
and neutral species at $\zabs=1.989$ toward HE\,2318$-$1107. Note that
 \MgI~$\lambda$2026 is partly blended with  \ZnII~$\lambda$2026, and
 so is \ZnII~$\lambda$2062 with \CrII~$\lambda$2062. \NI\ lines are blended
 altogether as well as with \lya\ forest absorptions but the column
 density can be derived in the central component.
%There are certainly
% additional weaker components on the very blue and very red side of
% the profile, that we would only detect by the \FeII\,$\lambda$1608 transition.
}
 \end{center}
\end{figure}

%The singly ionized metal components are numerous in this system, 
Ten highly blended components are needed to reproduce the whole absorption
profile of low ionization species that is spread over about 200
km~s$^{-1}$ (see Fig.~\ref{HE2318metals}). 
The profile is characterized by a central and strong component which is
the only one detected in H$_2$ and \CI.
\SII\ and \SiII\ are detected along the whole profile.
The decomposition is well constrained by the \FeII~$\lambda$1608 profile
that is redshifted in a region with high SNR and is nowhere saturated
except for the central component which is constrained by
the \FeII\,$\lambda\lambda2249,2260$ transition lines.
%, which decomposition into the ten components is grandly help by the
%\FeII~$\lambda$1608 transition. 
\CrII, \ZnII, \NiII\ and \MgI\ are detected in, respectively, seven, eight, eight and three
components. \NI\ is also detected over the whole profile, but because of
many blends with intervening Ly-$\alpha$ absorptions, we consider the
corresponding column densities as upper limits only, except for the central
component where the column density is derived thanks to the three detected \NI\ transitions. 
\MgII\ is also detected in the central component and
possibly also in component \#\,9 (see Table~\ref{HE2318_metalsT}). We nevertheless consider the \MgII\ column
density in component \#\,9 as an upper limit because of a possible blend with Ly-$\alpha$
forest absorptions. 
Upper limits at 3\,$\sigma$ are derived for non-detected
components. 

Some additional absorption is %possibly 
present in the blue and red wings of the \FeII~$\lambda$1608 profile. Unfortunately there is no
stronger \FeII\ absorption line in our spectrum to constrain the
column densities well. If we include these wings, the total \FeII\
column density is increased by about 7\%.
%10$^{13.75}$\,cm$^{-2}$ 

%We choose not to fit these components, seen only in \FeII$\lambda$1608, 
%but rather give a 3\,$\sigma$ upper limit derived thanks to \FeII$\lambda$2620:
%$\log N$(\FeII)~$\leq$13.30.
The result of Voigt-profile fitting is shown on Fig.~\ref{HE2318metals} and 
summarized in Table~\ref{HE2318_metalsT}.
%Using \ZnII\ as a reference for metallicity, 
Summing up the column densities measured in all components,
we derive a metallicity of [Zn/H$]=-0.85\pm0.06$. 
%It must be noted however that there is an uncertainty on the metal abundance
%due to the fact that H$_2$ and C~{\sc i} are detected only in the central component.
%One could argue that only this component should be associated to
%all the neutral hydrogen and then the metallicity would be three 
%times smaller: [Zn/H$]=-1.38\pm0.05$. However, this is probably an extreme
%assumption as some unknown amount of neutral hydrogen is associated
%to the strong satellites of the main component (see Fig.~\ref{HE2318metals}). 
%
%The low-ionization metal profile clearly features a central strong component
%(\#\,5 in Table~\ref{HE2318_metalsT}) which corresponds to the H$_2$ and \CI\ component.
%
%
\addtolength{\tabcolsep}{-1pt}
\begin{table*}
\caption{\label{HE2318_metalsT} 
Low-ionization metals column densities in the $\zabs=1.989$ DLA toward HE\,2318$-$1107.
%Voigt-profile
%  fitting results for low-ionization metal lines from the
%  $\zabs=1.989$ DLA toward HE\,2318$-$1107
}
\begin{center}

\begin{tabular}{c c}
\hline
\hline
\begin{tabular}{r l l l c c}

\# & $\zabs$  & Ion (X)     & $\log N$(X)$^{a,b}$    &    $b$ &  $\Delta v_{\rm X/{H_2}}$\\%_{\ion{C}{i}}}$ \\
   &          &             &                        &        &  $\Delta v_{\rm X/\ion{C}{i}}$\\
   &          &             &                &[km\,s$^{-1}$]&   [km\,s$^{-1}$]    \\

\hline

1  & 1.98807(8) & \NI         & $\leq$~14.23   & 8.9~$\pm$~2.0&  ---        \\
   &            & \MgI        & $<$~11.50       & '' 	     &   ---          \\
   &            & \MgII       &  blend          & ''         &             \\
   &            & \SiII       & 13.73~$\pm$~0.06  & ''	     &             \\
   &            & \SII        & 13.33~$\pm$~0.06  & ''	     &             \\
   &            & \CrII       & $<$~11.70       & ''         &             \\
   &            & \FeII       & 13.52~$\pm$~0.05  & ''	     &             \\
   &            & \NiII       & $<$~12.30       & ''	     &             \\
   &            & \ZnII       & $<$~10.80        & ''	     &             \\

2  & 1.98827(8) & \NI         & $\leq$~13.79    & 8.6~$\pm$~0.2&  ---        \\
   &            & \MgI        & $<$~11.50       & '' 	     &    ---         \\
   &            & \MgII       &  blend          & ''         &             \\
   &            & \SiII       & 14.05~$\pm$~0.02  & ''	     &             \\
   &            & \SII        & 13.75~$\pm$~0.02  & ''	     &             \\
   &            & \CrII       & 11.96~$\pm$~0.08  &            &             \\
   &            & \FeII       & 13.71~$\pm$~0.04  & ''	     &             \\
   &            & \NiII       & 12.58~$\pm$~0.05  & ''	     &             \\
   &            & \ZnII       & 11.18~$\pm$~0.12  & ''	     &             \\

3  & 1.98845(3) & \NI         & $\leq$~13.53    & 8.4~$\pm$~1.5&  ---        \\		     
   &            & \MgI        & $<$~11.50       & ''	     &    ---         \\
   &            & \MgII       &  blend          & ''         &             \\
   &            & \SiII       & 14.23~$\pm$~0.05  & ''	     &             \\
   &            & \SII        & 14.00~$\pm$~0.04  & ''	     &             \\
   &            & \CrII       & $<$~11.70       & ''         &             \\
   &            & \FeII       & 13.43~$\pm$~0.03  & ''	     &             \\
   &            & \NiII       & 12.63~$\pm$~0.04  & ''	     &             \\
   &            & \ZnII       & 11.53~$\pm$~0.05  & ''	     &             \\
							       	 					      				      					      
4  & 1.98865(0) & \NI         & $\leq$~13.62    &10.7~$\pm$~4.0&  ---        \\
   &            & \MgI        & $<$~11.50       & '' 	     &    ---         \\
   &            & \MgII       &  blend          & ''         &             \\
   &            & \SiII       & 14.62~$\pm$~0.05  & ''	     &             \\
   &            & \SII        & 14.43~$\pm$~0.05  & ''	     &             \\
   &            & \CrII       & 12.37~$\pm$~0.03  & ''         &             \\
   &            & \FeII       & 14.13~$\pm$~0.02  & ''	     &             \\
   &            & \NiII       & 13.11~$\pm$~0.02  & ''	     &             \\
   &            & \ZnII       & 11.72~$\pm$~0.04  & ''	     &             \\
					    		           	
5  & 1.98888(9) & \NI   & 14.55~$\pm$~0.02  & 4.2~$\pm$~0.3& $+$0.7      \\
   &          & \MgI          & 12.32~$\pm$~0.05  & ''	     &$-$0.4       \\
   &          & \MgII         & 14.94~$\pm$~0.04  & ''         &             \\
   &          & \SiII         & 14.72~$\pm$~0.01  & ''	     &             \\
   &          & \SII          & 14.50~$\pm$~0.04  & ''	     &             \\		      
   &          & \CrII         & 12.60~$\pm$~0.01  & ''         &             \\
   &          & \FeII         & 14.35~$\pm$~0.01  & ''	     &             \\
   &          & \NiII         & 13.18~$\pm$~0.01  & ''	     &             \\
   &          & \ZnII         & 11.97~$\pm$~0.01  & ''	     &             \\
     
\end{tabular} &

\begin{tabular}{r l l l c c}
\# & $\zabs$  & Ion (X)     & $\log N$(X)$^{a,b}$    &    $b$ &  $\Delta v_{\rm X/{H_2}}$\\%_{\ion{C}{i}}}$ \\
   &          &             &                        &                &  $\Delta v_{\rm X/\ion{C}{i}}$\\
   &          &             &                &[km\,s$^{-1}$]&  [km\,s$^{-1}$]    \\
\hline

6  & 1.98899(3) & \NI         & $\leq$~13.57    & 4.7~$\pm$~0.5&  ---        \\
   &          & \MgI          & $<$~11.50       & ''	     &    ---         \\
   &          & \MgII         & $<$~14.00       & ''         &             \\
   &          & \SiII         & 14.03~$\pm$~0.02  & ''	     &             \\
   &          & \SII          & 13.87~$\pm$~0.02  & ''	     &             \\
   &          & \CrII         & 11.94~$\pm$~0.06  & ''         &             \\
   &          & \FeII         & 13.55~$\pm$~0.05  & ''	     &             \\
   &          & \NiII         & 12.51~$\pm$~0.05  & ''	     &             \\
   &          & \ZnII         & 10.99~$\pm$~0.13  & ''	     &             \\

7  & 1.98918(3) & \NI         & $\leq$~14.09    & 9.5~$\pm$~1.6&  ---        \\
   &            & \MgI        & $<$~11.50       & ''	     &    ---         \\
   &            & \MgII       & $<$~14.00       & ''         &             \\
   &            & \SiII       & 14.31~$\pm$~0.02  & ''	     &             \\
   &            & \SII        & 14.18~$\pm$~0.02  & ''	     &             \\ 
   &            & \CrII       & 11.83~$\pm$~0.11  & ''         &             \\
   &            & \FeII       & 13.92~$\pm$~0.08  & ''	     &             \\
   &            & \NiII       & 12.90~$\pm$~0.02  & ''	     &             \\
   &            & \ZnII       & 11.63~$\pm$~0.04  & ''	     &             \\
						 	       	 					      				      					      
8  & 1.98942(1) & \NI         & $\leq$~14.01    & 10.5~$\pm$~1.8&  ---       \\
   &          & \MgI          & 11.80~$\pm$~0.13  & ''	     &     ---        \\
   &          & \MgII         & $<$~14.00       & ''         &             \\
   &          & \SiII         & 14.40~$\pm$~0.03  & ''	     &             \\
   &          & \SII          & 14.07~$\pm$~0.06  & ''	     &             \\
   &          & \CrII         & 12.17~$\pm$~0.05  & ''         &             \\
   &          & \FeII         & 14.03~$\pm$~0.03  & ''	     &             \\
   &          & \NiII         & 12.96~$\pm$~0.02  & ''	     &             \\
   &          & \ZnII         & 11.44~$\pm$~0.13  & ''	     &             \\

9  & 1.98956(4) & \NI         & $\leq$~13.39    & 1.6~$\pm$~1.0&  ---        \\	
   &          & \MgI          & $<$~11.50       & ''	     &    ---         \\ 
   &          & \MgII         & $\leq$~14.48    & ''         &             \\
   &          & \SiII         & 13.85~$\pm$~0.08  & ''	     &             \\
   &          & \SII          & 13.61~$\pm$~0.08  & ''	     &             \\			   
   &          & \CrII         & $<$~11.70       &            &             \\
   &          & \FeII         & 13.42~$\pm$~0.04  & ''	     &             \\
   &          & \NiII         & $<$~12.30       & ''	     &             \\
   &          & \ZnII         & $<$~10.80       & ''	     &             \\

10 & 1.98970(7) & \NI         & $\leq$~14.01    & 9.2~$\pm$~1.9& ---         \\
   &          & \MgI          & 12.04~$\pm$~0.08  & ''	     &   ---          \\
   &          & \MgII         &  blend          & ''         &             \\
   &          & \SiII         & 14.42~$\pm$~0.03  & ''	     &             \\
   &          & \SII          & 13.95~$\pm$~0.04  & ''	     &             \\
   &          & \CrII         & 12.26~$\pm$~0.04  & ''         &             \\
   &          & \FeII         & 13.99~$\pm$~0.08  & ''	     &             \\
   &          & \NiII         & 12.84~$\pm$~0.03  & ''	     &             \\
   &          & \ZnII         & 11.53~$\pm$~0.05  & ''	     &             \\

\end{tabular}\\
\hline
\end{tabular} 
\end{center}
\footnotesize
$^a$ 3\,$\sigma$ upper-limits are derived for non detected components, and marked by ``$<$''.\\
$^b$ Components for which the column density is uncertain because of
possible blends are considered as upper-limits, and marked by
``$\leq$''.
%$^c$ km\,s$^{-1}$.
\normalsize
\end{table*}
\addtolength{\tabcolsep}{1pt}

%\clearpage  %ONLY FOR REFEREE FORMAT

%%%%%%%%%%%%%%%%%%%%%%%%%%%%%%%%%%%%
%%% HE 0027
%%%%%%%%%%%%%%%%%%%%%%%%%%%%%%%%%%%%

\section{The system at $\zabs=2.402$ toward HE\,0027$-$1836}
\begin{figure}[!ht]
 \begin{center}
 \includegraphics[clip=,width=0.9\hsize]{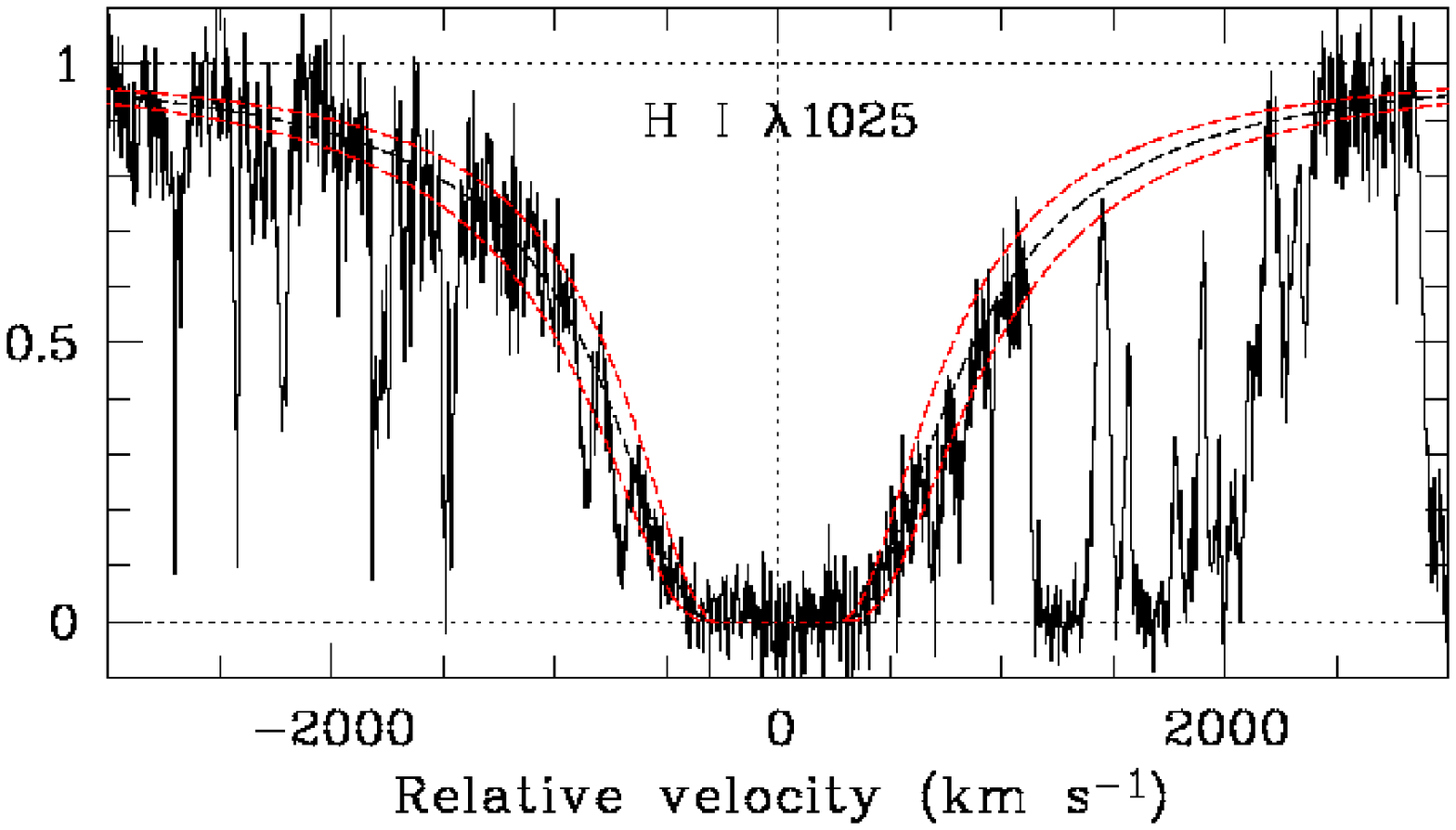}
 \caption{\label{HE0027_HI} Fit to the \lyb\ line at $\zabs=2.402$ toward
 HE\,0027$-$1836. The neutral hydrogen column density is $\log
 N$(\HI)~$=21.75\pm0.10$. {The dashed curves correspond to the
 best-fitted Voigt profile (black) and the associated errors (red).}}
 \end{center}
\end{figure}

The very broad \lya\ absorption line at $\zabs=2.402$ lies in the wing of the \lya\ emission line
from the quasar ($z_{\rm em}=2.56$) where the normalization of the spectrum 
is problematic. 
%From Voigt-profile
We therefore fitted a Voigt-profile to the \HI\ \lyb\ line (see Fig.~\ref{HE0027_HI})
to derive a total neutral hydrogen column density of $\log
N$(\HI)~=~$21.75\pm0.10$. 
This is one of the highest column densities measured in QSO-DLAs 
\citep[see,  e.g.][]{Prochaska05}.
%Three other absorption systems are identified along the line of sight, at
%$\zabs \simeq$~1.024; 2.421 and 2.469, thanks to
%\FeII\,$\lambda\lambda\lambda\lambda$2374,2382,2586,2600, and \MgII\,$\lambda\lambda$2796,2803 for
%the lowest redshift system, and \CIV\,$\lambda\lambda$1548,1550 for the two
%other ones. Note that the system at $\zabs \simeq 2.421$ is separated
%from the DLA system by only $\Delta v \sim 1700$~km\,s$^{-1}$. 
%The expected \lya-absorption would lie inside the
%\dla-absorption at $\zabs=2.042$. However, no associated
%low-ionization metals have been detected, the
%corresponding \HI-contribution to the \dla\  must be negligible and
%the fit is mainly constrained by \lyb\ and \lyg.
%

\subsection{Molecular content}
The wavelength range of the UVES spectrum allows us to observe a large number of H$_2$
absorption lines at the redshift of the DLA. A single H$_2$ component 
is detected in the rotational levels J~=~0 to J~=~5 with a possible detection of 
the J~=~6 rotational level. In addition to the overall normalization of the
spectrum, the continuum has been carefully normalized over about
100 to 800 km\,s$^{-1}$ locally around each line. 
It must be noted that this system is exceptional amongst H$_2$-bearing
DLAs, with more than 70 detected H$_2$ transitions. The absorption can be fitted with a single
component and most of the lines are not strongly saturated. 
In addition, at this redshift ($z_{\rm abs}$~$\sim$~2.4) blending is not
strong as the \lya\ forest is not dense and the data are of high signal-to-noise
ratio and high spectral resolution ($R\simeq50\,000$). Therefore we are in an ideal
position to derive with high confidence the column density $N$ and Doppler 
parameter $b({\rm J})$ for each rotational level J.
The fit was performed by $\chi^2$ minimization, considering different $b$-values for each
rotational level.
The relative optical depths of the different absorption 
lines with different oscillator strengths provide strong constraints on $N$ but 
more importantly on $b$. This is particularly true for rotational levels J~=~2 
and J~=~3 for which a large number of transitions with different oscillator strengths are 
observed. Some of the absorptions are in the intermediate regime where the 
equivalent widths strongly depend on $b$ (corresponding to the flat part of the 
curve-of-growth). 
Note that damping wings appear for some lines of the J~=~0 and J~=~1
rotational levels. %, providing another constraint on $b$. 
For the highest rotational levels (J~=~4 and J~=~5), absorption lines are optically thin, 
making $b$ slightly less constrained than for lower rotational levels. However,
it is apparent that these lines are broader than the J~=~2 and J~=~3 features.
%
%the broadening is
%directly apparent, although the corresponding absorption lines are optically thin, 
%making $b$ slightly less constrained than for rotational levels J~$\le$~3.
Two consistent absorption features are detected for rotational level J~=~6, but the
lines are too weak to perform a fit with an independent $b$. We
therefore have chosen to fix $b$(J~=~6) equal to $b$(J~=~5) previously measured independently. 
The derived J~=~6 column density should be preferably considered as an
upper limit due to possible blends (see Table~\ref{HE0027H2CI}).
The results of the Voigt-profile fitting are shown in 
Figs.~\ref{HE0027_J0} to \ref{HE0027_J6} and summarized in
Table~\ref{HE0027H2CI}. The molecular fraction in this system is $\log
f = -4.15\pm0.17$.
The important finding here is that the Doppler parameter $b$ increases 
with J.

\addtolength{\tabcolsep}{-1pt}
\begin{table}[!ht]
\caption{\label{HE0027H2CI} 
H$_2$ and \CI\ column densities in the $\zabs=2.402$ DLA toward HE\,0027$-$1836.
%Voigt-profile fitting results for different
%  H$_2$ rotational levels and \CI\ ground-state fine-structure levels
%  in the $\zabs=2.402$ \dla\ system toward HE\,0027$-$1836.
}
 \begin{center}
  \begin{tabular}{l l l l c c}
\hline
\hline
\#$^a$ & $\zabs$  & level & $\log N($H$_2$, J)\,$^{b}$ & $b$           & $T_{\rm 0-J}$\,$^{c}$\\
   &          &       &                 &[km\,s$^{-1}$]& [K]     \\

\hline
3  & 2.40183(4) & J~=~0 & 16.75~$\pm$~0.06  & 1.06~$\pm$~0.13 & ---               \\
   &            & J~=~1 & 17.15~$\pm$~0.07  & 1.46~$\pm$~0.14 & 134$_{-55}^{+317}$ \\
   &            & J~=~2 & 14.91~$\pm$~0.02  & 2.67~$\pm$~0.08 &  88$_{-8}^{+9}$  \\
   &            & J~=~3 & 14.91~$\pm$~0.01  & 3.77~$\pm$~0.07 & 141$_{-9}^{+10}$  \\
   &            & J~=~4 & 14.22~$\pm$~0.01  & 4.61~$\pm$~0.25 & 213$_{-12}^{+14}$  \\
   &            & J~=~5 & 14.02~$\pm$~0.03  & 6.17~$\pm$~0.69 & 262$_{-16}^{+18}$  \\
   &            & J~=~6 & $\leq$~13.53      & 6.17$^{d}$& $\leq 375$  \\
\hline
   &            &       &                 &               &                    \\
\hline
\hline
\#$^a$ & $\zabs$    & $\log N$(\CI)  & $\log N$(\CI$^*$)    &    b                &$\Delta v_{\ion{C}{i}/{\rm H}_2}$ \\
   &            &                &                      &[km\,s$^{-1}$]       & [km\,s$^{-1}$]  \\
\hline										
 3 & 2.40185(8) &  12.25$_{-0.15}^{+0.09}$ & $\leq$~12.27 &   0.6-3.0
 &  $+$2.1            \\
\hline
\end{tabular}
 \end{center}
\footnotesize
$^a$ The numbering refers to that of the metal components (see Table~\ref{HE0027_metalsT}).\\
$^{b}$  The error on the H$_2$ column densities are the
1\,$\sigma$ uncertainty from fitting the Voigt-profiles.\\
$^{c}$ The excitation temperatures are calculated using the
3\,$\sigma$ range for $\log N$(H$_2$,J).\\
$^{d}$ The Doppler parameter for J~=~6 has been fixed to be equal
to $b$(J~=~5) derived independently.
\normalsize
\end{table}
\addtolength{\tabcolsep}{1pt}

\subsection{Neutral carbon}
Neutral carbon is detected in one component at about the same redshift as the H$_2$
component. The absorption line is optically thin, and the
column density therefore does not depend much on $b$.
%, that is not well constrained.
The corresponding \CI$^*$ absorption is weak and detected just above the 3\,$\sigma$
confidence level ($\log N$(\CI$^*$)~=~12.2). The derived column density is thus
uncertain and should be considered as an upper limit: $\log
N$(\CI$^*$)~$\le$~12.27 (see Table~\ref{HE0027H2CI}).
%Though close to the 1\,$\sigma$ error (about 1~km\,s$^{-1}$), 
We observe a small velocity shift between the positions of the \CI\ and H$_2$ 
components %This is probably within errors however. 
but no velocity shift between \CI\ and component \#3 of the
low-ionization metal profile.
%, confirming $\lambda$(\CI) despite the weakness, poor SNR, and small blends of \CI\ lines.
The upper limit on $N$(\CI$^*$)$/N$(\CI) leads to an
upper limit on the neutral hydrogen density, $n_{\rm H} <
150$~cm$^{-3}$, for a typical temperature of 100~$< T <$~200~K and a
Galactic UV radiation field \citep[see
  also][]{Noterdaeme07}.

\begin{figure}[!ht]
 \begin{center}
 \includegraphics[clip=,width=0.92\hsize]{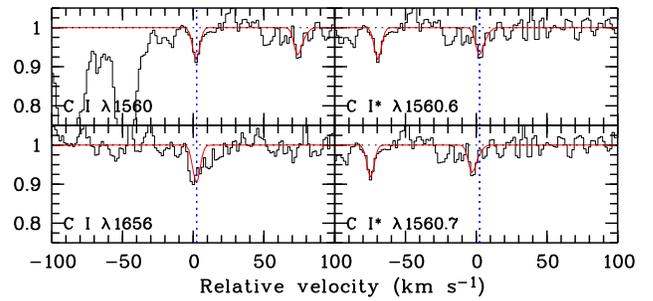}
 \caption{Neutral carbon absorption lines %at $\zabs=2.40186$
 toward HE\,0027$-$1836. 
%{\bf The origin ($v=0$~km\,s$^{-1}$) of the
% velocity scale corresponds to $z=2.40183$.} 
The \CI$^*$ absorption is weak and 
%close to the noise level. One should therefore 
the derived column density is thus considered as an upper limit. \label{HE0027_CI} }
 \end{center}
\end{figure}

\subsection{Metal content}

%Three components are used to fit the metal profile of this system.
The profile is made of two strong well separated features ($\Delta v\sim25$~km\,s$^{-1}$), 
the first one being a blend of two components. The second feature, 
corresponding to the third and also the strongest component in the profile , 
is associated with the H$_2$ component but is significantly broader.
We detect and measure column densities for a large number of species,
namely \NI, \MgI, \MgII, \SiII, \PII, \SII, \ArI, \TiII, \CrII, \MnII,
\FeII, \NiII, and \ZnII. In addition, we measure 3\,$\sigma$ upper limits on the column densities of
\PbII\ (resp. \CuII) for each component: $\log N($\PbII$)< 11.1$
(respectively, $\log N($\CuII$)<12.1$) from the non-detection of \PbII~$\lambda$1433
at SNR~$\sim$~46 (resp. \CuII~$\lambda$1358; SNR~$\sim$~16).
%Because the blending of the two first components is severe, the decomposition 
%between the two components is degenerated and it is
%therefore better to consider the whole feature (components 1 and 2),
%for which the total column density is more accurate.
\par
\begin{figure*}[!ht]
 \begin{center}
\begin{tabular}{cc}
 \includegraphics[clip=, width=0.48\hsize]{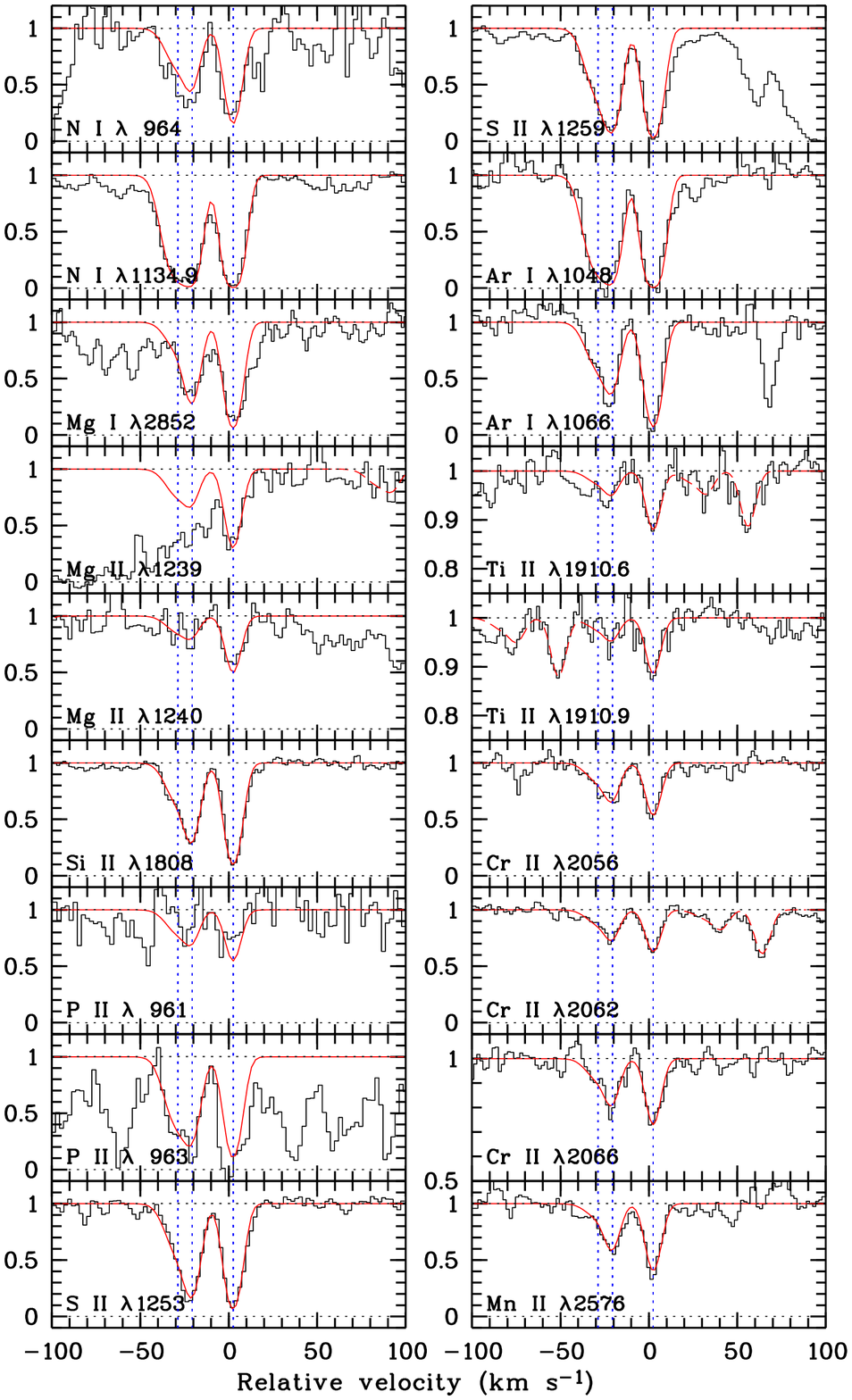} &
 \includegraphics[clip=,width=0.48\hsize]{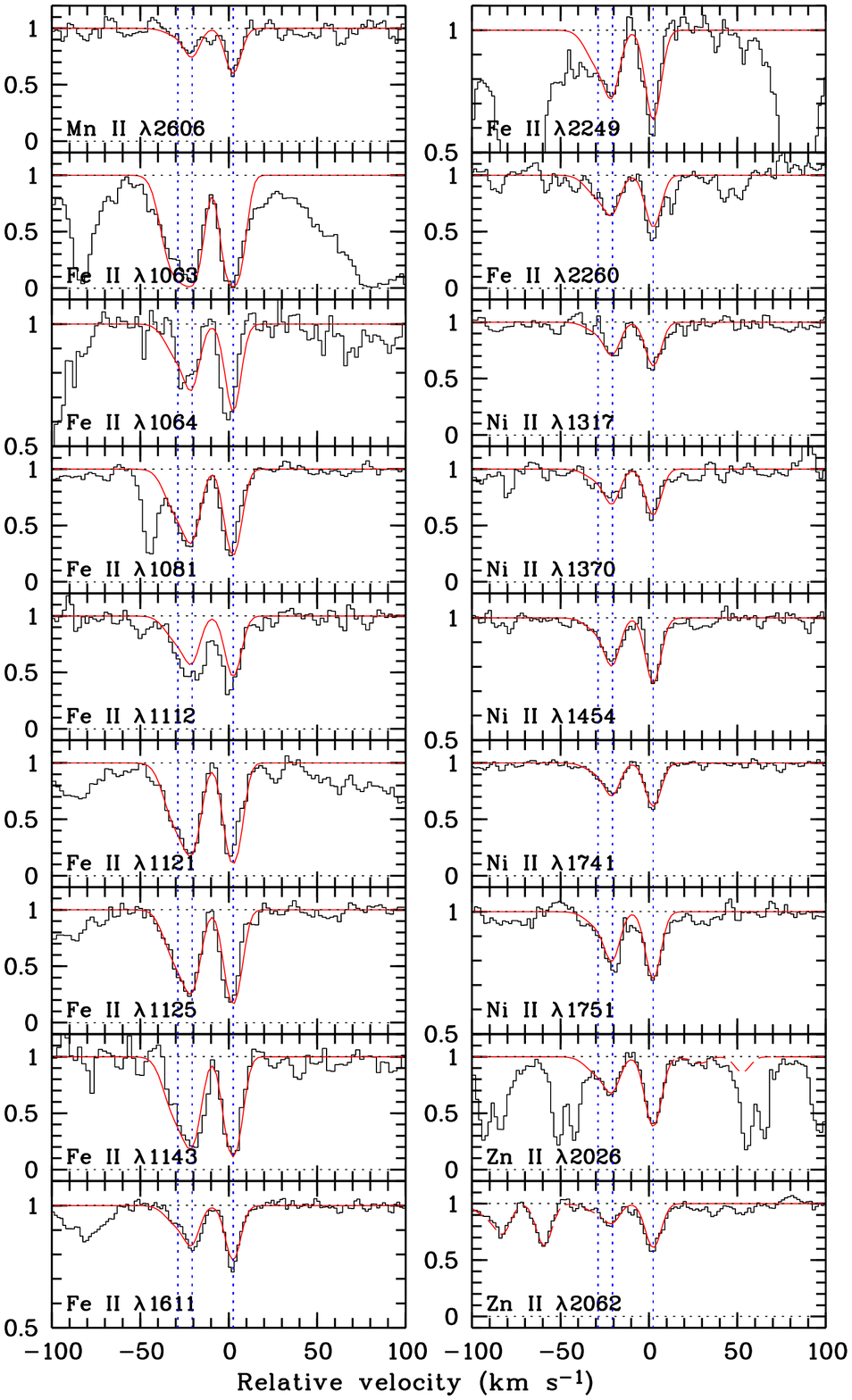} \\
\end{tabular}
 \caption{\label{HE0027_metals} Absorption profiles of low-ionization
 and neutral species from the
 $\zabs=2.402$ DLA system toward HE\,0027$-$1836. 
%The origin ($v=0$~km\,s$^{-1}$) of the
% velocity scale corresponds to the H$_2$ component. 
 The synthetic best-fit spectrum is
 superimposed to the observed spectrum. The dashed lines in some panels
 mark absorption from another element. For example, there is some
 \MgI\,$\lambda$2026 absorption near the \ZnII\,$\lambda$2026 one. The
 transitions \FeII\,$\lambda\lambda\lambda$1063,1064,1112 are shown but were not
 used during the fitting process because of uncertainties in atomic parameters.}
 \end{center}
\end{figure*}

 To derive the characteristics of the lines, we first determined the $b$-values 
by fitting unblended and good SNR lines. We then fixed the
obtained $b$ values and added all other detected transitions.
Although relative wavelengths were relaxed when fitting the lines, 
%there is no
the final fit shows no velocity-shift between the \CI\ component and the third metal 
component.
%A small shift is seen between \CI\ and H$_2$, but it is within errors.
Note that the best oscillator strength found in the literature for
\FeII~$\lambda$1112, $f=0.00629$ \citep{Howk00}, seems to be underestimated
(see Fig.~\ref{HE0027_metals}) unless there is extra absorption (possibly from 
the \lya\ forest) at that wavelength. The oscillator strength from
\citet{Morton03} is even smaller, $f=0.00446$. 
%commented referee.
%It also appears that the wavelength of \SII\,$\lambda$1253 reported by
%\citet{Morton03}, $\lambda_{\rm vac}=1253.805$, is not accurate. 
{Using the measured $\lambda_{\rm vac}=1253.805$ for \SII\,$\lambda$1253
  \citep{Morton03}} % this value
induces a small but significant shift between the observed and the synthetic
profiles. To make the two spectra consistent the {Ritz} wavelength 
$\lambda_{\rm vac}=1253.811$ \citep{Morton91} had to be used instead. 
%The latter wavelength should therefore be used.
The metallicity derived for this system is [Zn/H$]=-1.63\pm0.10$.
A study of the highly ionized species is presented by \citet{Fox07a,Fox07b}.

\begin{table}
\caption{\label{HE0027_metalsT} 
Low-ionization metals column densities in the $\zabs=2.402$ DLA toward HE\,0027$-$1836.
%Voigt-profile fitting results for
%  low-ionization metal lines from the $\zabs=2.402$ DLA toward
%  HE\,0027$-$1836.
}
\begin{center}
\begin{tabular}{r l l l c c}
\hline
\hline
\# & $\zabs$    & Ion (X)     & $\log N$(X)    &    $b$&$\Delta v_{\rm X/{H_2}}$\\%_{\ion{C}{i}}}$\\
  &            &             &                &                  &$\Delta v_{\rm X/\ion{C}{i}}$\\
   &            &             &                &[km\,s$^{-1}$]  &   [km\,s$^{-1}$]          \\
\hline

1  & 2.40150(4) & \NI\        & 14.53~$\pm$~0.24 & 8.4~$\pm$~1.5    & --- \\
   &            & \MgI\       & 11.52~$\pm$~0.09 & ''             &---\\
   &            & \MgII\      & 15.32~$\pm$~0.14 & ''             &\\
   &            & \SiII\      & 14.84~$\pm$~0.11 & ''             &\\
   &            & \PII\       & 12.62~$\pm$~0.30 & ''             &\\
   &            & \SII\       & 14.41~$\pm$~0.10 & ''             &\\
   &            & \ArI\       & 13.64~$\pm$~0.06 & ''             &\\
   &            & \TiII\      & 11.87~$\pm$~0.29 & ''             &\\
   &            & \CrII\      & 12.71~$\pm$~0.04 & ''             &\\
   &            & \MnII\      & 11.93~$\pm$~0.16 & ''             &\\
   &            & \FeII\      & 14.36~$\pm$~0.05 & ''             &\\
   &            & \NiII\      & 12.88~$\pm$~0.10 & ''             &\\
   &            & \ZnII\      & 11.99~$\pm$~0.10 & ''             &\\

2  & 2.40159(6) & \NI\        & 14.51~$\pm$~0.22 & 4.7~$\pm$~0.9    & --- \\
   &            & \MgI\       & 11.96~$\pm$~0.09 & ''             &---\\
   &            & \MgII\      & 15.14~$\pm$~0.17 & ''             &\\
   &            & \SiII\      & 15.06~$\pm$~0.07 & ''             &\\
   &            & \PII\       & 12.50~$\pm$~0.30 & ''             &\\
   &            & \SII\       & 14.68~$\pm$~0.03 & ''             &\\
   &            & \ArI\       & 13.64~$\pm$~0.06 & ''             &\\
   &            & \TiII\      & 11.90~$\pm$~0.11 & ''             &\\
   &            & \CrII\      & 12.81~$\pm$~0.04 & ''             &\\
   &            & \MnII\      & 12.31~$\pm$~0.08 & ''             &\\
   &            & \FeII\      & 14.39~$\pm$~0.05 & ''             &\\
   &            & \NiII\      & 13.21~$\pm$~0.05 & ''             &\\
   &            & \ZnII\      & 12.09~$\pm$~0.06 & ''             &\\

3  & 2.40185(8) & \NI\        & 15.04~$\pm$~0.09 & 4.7~$\pm$~0.2    & $+$2.1 \\
   &            & \MgI\       & 12.41~$\pm$~0.05 & ''             &   0.0\\
   &            & \MgII\      & 15.81~$\pm$~0.02 & ''             &\\
   &            & \SiII\      & 15.45~$\pm$~0.02 & ''             &\\
   &            & \PII\       & 12.87~$\pm$~0.40 & ''             &\\
   &            & \SII\       & 14.98~$\pm$~0.03 & ''             &\\
   &            & \ArI\       & 14.24~$\pm$~0.02 & ''             &\\
   &            & \TiII\      & 12.41~$\pm$~0.03 & ''             &\\
   &            & \CrII\      & 13.08~$\pm$~0.01 & ''             &\\
   &            & \MnII\      & 12.60~$\pm$~0.02 & ''             &\\
   &            & \FeII\      & 14.66~$\pm$~0.03 & ''             &\\
   &            & \NiII\      & 13.42~$\pm$~0.02 & ''             &\\
   &            & \ZnII\      & 12.60~$\pm$~0.01 & ''             &\\

\hline
\end{tabular}
\end{center}
\footnotesize
%$^a$ km\,s$^{-1}$.
\normalsize
\end{table}

\subsection{\CII$^*$ and the ambient UV flux \label{secCII}}
\begin{figure}[!ht]
 \begin{center}
 \includegraphics[clip=,width=0.98\hsize]{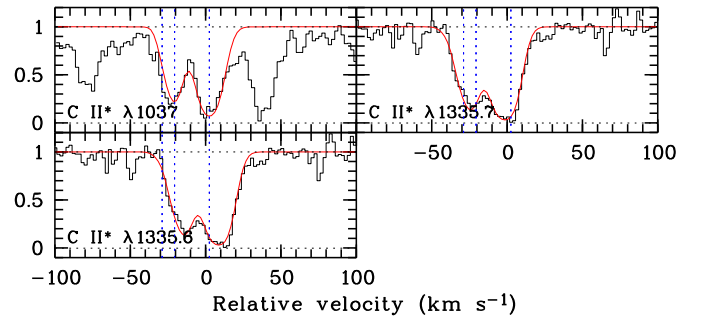}
 \caption{\CII$^*$ absorption profile in the DLA system toward
 HE\,0027$-$1836. The absorption seen at $v \simeq +35$~km\,s$^{-1}$ in
 the first panel is due to the H$_2$\,L5--0\,R1 transition. \label{HE0027_CIIstar} }
 \end{center}
\end{figure}

Strong and saturated C~{\sc ii}$^*$$\lambda$1335 absorption 
%by excited singly ionized carbon
%\CII$^*$ ($\lambda_{\rm rest}\simeq1335 {\rm \AA}$) 
is seen in this system (see Fig.~\ref{HE0027_CIIstar}).
%The feature with $\lambda_{\rm rest}\simeq1335 {\rm \AA}$ is strong and
%saturated. However 
We can use however the weaker transition at
$\lambda_{\rm rest}\simeq1037 {\rm \AA}$ to derive reliable column
density estimates (see Table~\ref{HE0027_CIIstart}). 
%with intervening H~{\sc i} Ly-$\alpha$ absorptions. 
%To allow fitting of the lines, 
During the fit redshifts of the \CII$^*$ components were constrained
to be the same as those of other low-ionization species.
In turn, Doppler parameters could not be given the same values. 
%reproduce the \CII$^*$ absorption features. 
In any case the $b$-values for \CII$^*$ are not well constrained, especially
for the weak component \#\,1 but this does not strongly affect the derived
column densities. The associated uncertainties were derived by
considering a range of $b$ values. A large Doppler parameter is needed for component 
\#\,3. This probably reveals the multiple component nature of the feature.
%This may be due to an additional component that would be revealed at such 
%large optical depth. Errors given in Table~\ref{HE0027_CIIstart} take
%this possible effect into account.
Following \citet{Wolfe03}, we can estimate the UV ambient
flux in the system, considering the equilibrium between photo-electric
heating and cooling by \CII\,158\,$\mu$m emission.
The heating of the gas by photo-electric effect on dust grains is
proportional to the dust-to-gas ratio, ($\kappa = 10^{\left[{\rm
      Zn/H}\right]} (1-10^{\left[{\rm Fe/Zn}\right]})$), and to the UV
flux, $F_{\rm UV}$. 
The cooling rate is proportional to the \CII\,158\,$\mu$m emission per H
atom, as estimated by \citet{Pottasch79}:
\begin{equation}
l_c = {{N(\ion{C}{ii}^*) h \nu_{ul} A_{ul}}\over {N(\ion{H}{i})}}~{\rm
  erg}\,{\rm s}^{-1}~{\rm per~H~atom,}
\end{equation}
where $A_{ul}$ is the coefficient for spontaneous emission of the
$^2P_{3/2} \rightarrow$ $^2P_{1/2}$ transition and $h \nu_{ul}$ the
corresponding energy. We find $\log l_c \simeq -27 \pm 0.2$.
% for $\kappa \simeq 0.02$.
%The equilibrium between cooling and heating of the gas
The conditions in the gas can be compared to that in the Galactic ISM, 
considering the same type of dust grains for both media:
\begin{equation}
{{l_c}\over{l_c^{gal}}} \approx {{\kappa}}\,{{F_{UV}}\over{F_{UV}^{gal}}}
\end{equation}
\par\noindent From Fig.~2 of \citet{Wolfe03} \citep[see also][]{Lehner04} it can be seen that 
$\log l_c^{gal} \approx -26.6$ in the Galaxy at $N(\ion{H}{i}) = 10^{21.75}$~cm$^{-2}$.
We then estimate ${F_{UV}}/{F_{UV}^{gal}}\sim 20$ for $\kappa \simeq 0.02$.
The UV flux in the DLA is about an order of magnitude higher than in
the Galactic disk. Note that we ignored the dependence of the photo-electric
heating on the temperature and density of the gas
\citep{Weingartner01}, and that the result is therefore only valid for
temperatures and densities similar to that in our Galaxy. 

\begin{table}
\caption{\label{HE0027_CIIstart} %Voigt-profile fitting results for
  \CII$^*$ column densities in the $\zabs=2.402$ DLA toward HE\,0027$-$1836.}
\begin{center}
\begin{tabular}{l l l c}
\hline
\hline
\#$^a$ & $\zabs$    & $\log N($\CII$^*)$ & $b$     \\
   &            &                    & [km\,s$^{-1}$]\\
\hline
1  & 2.40150(4) & 12.65~$\pm$~0.30       & 3-8 \\
2  & 2.40159(6) & 13.80~$\pm$~0.10       & 4-7 \\
3  & 2.40185(8) & 14.15~$\pm$~0.25       & 7-9 \\
\hline
\end{tabular}
\end{center}
\footnotesize
$^a$ The numbering refers to that of the low-ionization metal
components (see Table~\ref{HE0027_metalsT}).
\normalsize
\end{table}

\section{Excitation of H$_2$ toward HE\,0027$-$1836}
In this Section, we analyze in detail the excitation of H$_2$ at
$\zabs=2.40183$ toward
HE\,0027$-$1836 where absorptions from rotational levels J~=~1 to
J~=~5 and possibly J~=~6 are seen in a single component.

% which is the only system presented with detected high
%rotational levels. Furthermore, the quality of the data allows a
%detailed study of the H$_2$ lines.

\subsection{Excitation temperatures \label{secH2ex}}
\begin{figure}[!ht]
 \begin{center}
 \includegraphics[clip=,width=\hsize]{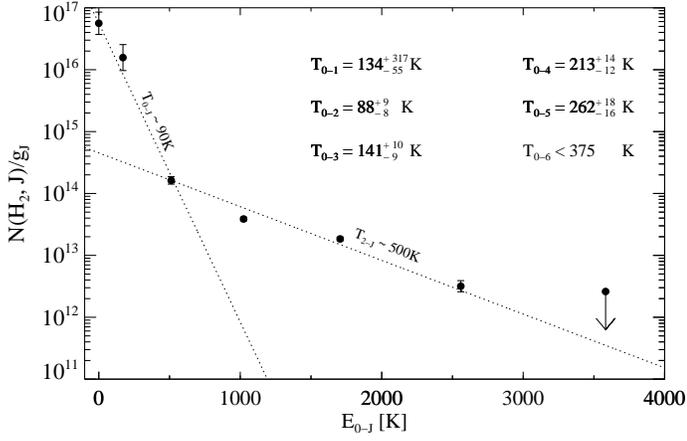}
 \caption{Excitation diagram of H$_2$ at
 $\zabs=2.40183$ toward HE\,0027$-$1836. It is apparent that it
is characterized by two excitation temperatures: $T_{\rm 0-J} \sim
 90$~K for J~$\leq$~2 and $T_{\rm 2-J} \sim 500$~K for J~$>$~2.  
%The population of the J~$>$~2 levels clearly do not follow a 
%Boltzman distribution. 
\label{HE0027H2_tex} }
 \end{center}
\end{figure}

Figure~\ref{HE0027H2_tex} shows the H$_2$ excitation diagram in this
system. The graph gives for the different rotational levels
$N$(H$_2$,J)/g$_{\rm J}$, where g$_{\rm J}$ is the statistical weight
of the level, versus the relative energy between that level and J~=~0. 
The slope $s$ of a straight line in the graph is inversely proportional to the excitation
temperature, $s=-1/\left({T_{\rm J-J^{\prime}} \ln{10}}\right)$.
The kinetic temperature is generally estimated by $T_{0-1}$ assuming
the J~=~1 level is thermalised. Here, $T_{0-1} \simeq 134$~K.
%We will show in the next Section that if the Doppler parameter is not due
%to macroscopic movements (turbulence), the effective kinetic
%temperature of individual rotational levels is actually an increasing function of J.

\noindent The H$_2$ excitation diagram is characterized by two excitation
temperatures, $T_{\rm 0-J} \sim 90$~K for J~$\leq$~2 and $T_{\rm 2-J} \sim 500$~K
 for J~$>$~2 (see Fig.~\ref{HE0027H2_tex}).
%The rotational levels J~$>$~2 are not Boltzman-distributed. 
Therefore, other processes than collisions at a temperature of the order of $T\sim 100$~K
(the approximate kinetic temperature) are at play. 
%Those are generally thought to be 
Formation pumping (just after formation, H$_2$ molecules are in high-J levels) and UV
pumping from low-J levels upwards are likely to be the most important of these processes. 
Assuming these are the main
processes allows us to estimate the strength of the surrounding 
UV radiation field. 
Writing the equilibrium between the processes populating and depopulating
the J~=~4 rotational level $[$see Eq.~5 in \citet{Noterdaeme07}, also \citet{Hirashita05, Cui05}$]$, 
we can estimate the photo-absorption
rate in the J~=~0 level, $\beta_0$, inside the H$_2$-bearing cloud: $\beta_0 \simeq
2.5\times10^{-11}$\,s$^{-1}$. This is a very low value, 
%showing that the UV excitation of H$_2$ inside the cloud is
%negligible. This is 
probably due to self- and dust-shielding by the outer layers of the cloud.
The corrective term for shielding is expressed as \citep{Draine96}:

\begin{equation}
S_{\rm shield}=\left({N({\rm H}_2) \over {\rm 10^{14}\,cm^{-2}}}\right)^{-0.75} e^{-\tau_{\rm UV}}
\label{shield}
\end{equation}

\par \noindent This is the product of two terms, the first term, $S_{\rm H_2}$, is due to 
self-shielding, the second, $S_{\rm dust}$, is due to dust extinction. The total H$_2$ 
column density in the cloud is $N($H$_2) = 10^{17.3}$\,cm$^{-2}$, leading to 
$S_{\rm H_2} \simeq 0.00335$.  Dust extinction is calculated
using the dust optical depth, $\tau_{\rm UV}$:%\equiv\sigma_{\rm d} N_{\rm d}$:

\begin{equation}
\tau_{\rm UV}= 0.879\,{\left({a \over {\rm 0.1 \mu m}}\right)}^{-1} \left({\delta \over {\rm 2\,g\,cm^{-3}}}\right)^{-1} \left({g\over{10^{-2}}}\right) \left({N_{\rm H} \over
      {\rm 10^{21}\,cm^{-2}}}\right)
\label{dust_shield}
\end{equation}
\par \noindent where $a$ is the radius of a grain, $g$ is the dust-to-gas
mass ratio and $\delta$ is the grain material density \citep{Hirashita05}.
This optical depth is difficult to estimate, due to the unknown type of dust
and the uncertainty on the exact \HI\ column density in the H$_2$-bearing component. 
We can assume $a=0.1$~${\rm \mu}$m, $\delta=2$~g\,cm$^{-3}$, and scale the dust-to-gas
mass ratio $g$ with the dust-to-gas ratio $\kappa=10^{\rm [Zn/H]}\left({1-10^{[\rm Fe/Zn]}}\right)$
\citep{Prochaska_W02}.
We then get:
\begin{equation}
\tau_{\rm UV} = 0.879\,\kappa\,\left({{N_{\rm H}} \over {10^{21}\,{\rm cm}^{-2}}}\right)
\end{equation}

\par \noindent or equivalently,

\begin{equation}
\tau_{\rm UV} = 0.879 \left({{N({\rm Zn})}\over{10^{\rm [Zn/H]_\odot}}} -
    {{N({\rm Fe})}\over{10^{\rm [Fe/H]_\odot}}}\right) \left(1 \over
      {\rm 10^{21}\,cm^{-2}}\right)
\end{equation}

\par \noindent where the column densities are those measured in the H$_2$-bearing component.
%Assuming that all the \HI\ is associated to the H$_2$ bearing component, 
We obtain $S_{\rm dust} = 0.94$ and $S_{\rm shield} = S_{\rm
  H_2} S_{\rm dust} \simeq 0.003$. Note that the self-shielding
dominates the total shielding.
%, and that dust-shielding only  contributes to about 6\%. 
We can then estimate the UV flux outside the cloud \citep[see][]{Noterdaeme07}:
\begin{equation}
 J_{\rm LW} \simeq {{8 \times 10^{-11} \beta_0} \over { S_{\rm
 shield}}} \simeq 6.3 \times 10^{-19}\,{\rm erg}\,{\rm s}^{-1}\,{\rm
 cm}^{-2}\,{\rm Hz}^{-1}\,{\rm sr}^{-1}.
\label{J_beta_S}
\end{equation}

\par \noindent This is about 20 times higher than in the solar vicinity \citep[$J_{\rm
    LW,\odot}\simeq 3.2 \times
  10^{-20}$\,erg\,s$^{-1}$\,cm$^{-2}$\,Hz$^{-1}$\,sr$^{-1}$;][]{Habing68}.
Again, all this assumes that the high rotational levels of
H$_2$ are populated by formation- and UV-pumping. If collisional
excitation, in case of turbulence or shocks for example, is playing an important 
role, then the derived UV flux should be considered as an upper limit. 
%The observed velocity dispersion of
%high-J levels actually emphasize the possibility of such
%excitations. 
%It is however reassuring that the UV flux estimated using \CII$^*$ is
%very close to the one using H$_2$. The two methods leads both to an ambient 
%UV flux of about twenty times the Galactic one.

\subsection{Velocity dispersion}
\begin{figure}[!ht]
 \begin{center}
 \includegraphics[width=\hsize]{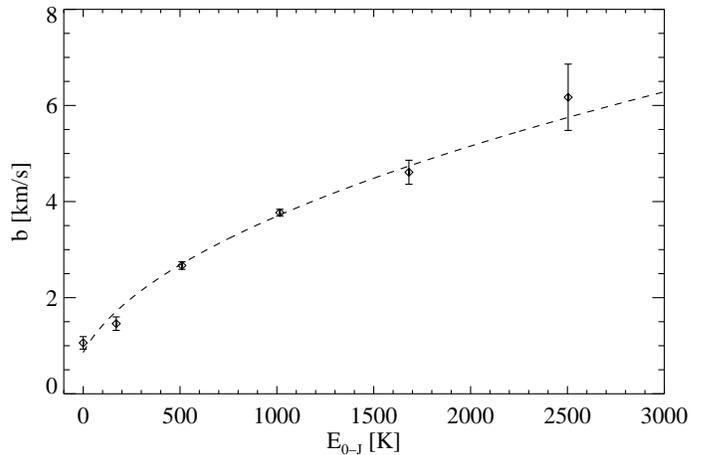}
 \caption{The Doppler parameter $b$ as a function of the energy
 between the rotational levels (from J~=~0 to J~=~5) in the DLA system toward
 HE\,0027$-$1836. The smallest errors are found for J~=~2 and 3, for which numerous lines with good SNR and
 in different optical depth regimes are observed. The dotted curve
 corresponds to a fit with a single free parameter (Eq.~\ref{bjeq}). \label{HE0027_b} }
 \end{center}
\end{figure}

In Fig.~\ref{HE0027_b}, we plot the Doppler parameter of the lines from a
given rotational level, $b$, 
as a function of the energy of the rotational level J.
It is apparent that higher J levels have broader lines. 
This effect has already been observed in the local interstellar  medium in some 
cases. It has first been derived from  curve-of-growth analysis
%  and the line widths 
in several {\it Copernicus} observations
  \citep{Spitzer73, Spitzer74}, then observed directly by
  \citet{Jenkins97} with R~$=$~120\,000 data from the Interstellar Medium
  Absorption Profile Spectrograph (IMAPS). In that case however
the increase in the line broadening is associated with a regular shift 
in velocity. More recently, the same effect has been reported by \citet{Lacour05} 
%showed evidence for the velocity dispersion of high rotational levels
%  H$_2$-lines increasing with the level of  excitation 
along four lines of sight toward early-type Galactic stars. 
In high redshift DLAs, \citet{Ledoux03} already suggested that the lines
of J~$\ge 2$ rotational levels require higher $b$ values than the lower 
J-level lines.
% in the high-redshift system toward Q\,0405$-$443.
  It is however the first time that a systematic effect is seen beyond any doubt
at high redshift.
\par\noindent
In case the broadening is only thermal, the Doppler parameter is
related to the kinetic energy by:
 \begin{equation}
  b({\rm J}) = \sqrt {2k_B E_k({\rm J})/(3m_{\rm H})}
 \end{equation}
\par\noindent
If we assume that the excess of kinetic energy of molecules in some J level 
compared to the kinetic energy of molecules in the J~=~0 level
is directly proportional to the energy of
  the rotational level, i.e. $E_k({\rm J}) - E_k(0) = a\,E_{\rm 0-J}$, with $a$
a constant, then we can try to fit the observed trend with the expression:

\begin{equation}
 b({\rm J}) = \sqrt{ 2k_B (E_k(0) + a\,E_{\rm 0-J})/\left({3 m_{\rm H}}\right)}
\label{bjeq}
\end{equation} 
%macs su
\par \noindent where $E_k(0)$ is the kinetic energy of rotational level J~=~0, for which we
use the excitation temperature $T_{0-1} \simeq 134$~K previously determined. 
%as an estimate.
% We can then perform a fit on $b(E_{\rm J})$ with a unique parameter. 
The least square minimization gives $E_k({\rm J})\simeq E_k(0) + 2.34\,E_{\rm 0-J}$. 
The result is shown on Fig.~\ref{HE0027_b}. One can see that
this very simple assumption fits the data very well. 
It is not easy to find an explanation to this effect.
\citet{Jenkins97} argued that such trend in $b$ seen along the line of sight
to $\chi$OriA could be explained if the gas is located behind a
J-shock. However, the column densities along the present line of sight 
are much larger and, more importantly, this would produce coherent velocity
shifts between the lines from different J levels, which we do not observe.

UV pumping cannot produce different $b$ values because the
  cascade following the UV absorption releases the energy through
  infrared photons, and such process does not result in any change in
  the molecule's kinetic energy.
%  Different possible mechanisms exists for
%  broadening the H$_2$ lines: turbulences, shocks and H$_2$-formation pumping.

%The broadening is related to the kinetic energy by:
% \begin{equation}
%  b({\rm J}) = \sqrt {2 E_k({\rm J})/(3m_{\rm H})}
% \end{equation} 
%If we assume that the gain of energy is directly proportional to the energy of
%  the rotational level (i.e. $E_k(J) = E_k(0) + a E_{\rm J}$, with $a$
%  constant), then we have the relation, expressing the energies in K:
%\begin{equation}
% b({\rm J}) = \sqrt{ 2k_B (E_k(0) + a E_{\rm J})/(3 m_{\rm H})}
%\label{bjeq}
%\end{equation} 
%$E_k(0)$ is the kinetic energy of rotational level J~=~0, for which we
%therefore use the excitation temperature $T_{0-1} \simeq
%134$~K previously determined as an estimate.

%We can then perform a fit on $b(E_{\rm J})$ with a unique parameter. The least
%square minimization gives $E_k({\rm J})\simeq E_k(0) + 2.34 E_{\rm
%  J}$. This result is shown on Fig.~\ref{HE0027_b}. One can see that
%the very simple assumption we made fits very well the data. 

\citet{Lacour05} show that such behavior can be explained by the
consequences of H$_2$-formation onto dust grains. 
After formation, the molecules escape with an excess kinetic energy 
left-over from the formation process.
The highest rotational levels have a radiative life-time
considerably shorter than the lowest J levels. This implies that 
low-J molecules have more time than high-J molecules to cool down through
collisions with \HI\ \citep[see, e.g.,][]{Spitzer73, Lacour05}. This implies 
that, if other processes are negligible, the kinetic energy of high-J
molecules could be larger than that of low-J ones. 
%This can explain also the non-Boltzman
%distribution of the J~$>2$ levels as seen in Fig.~\ref{HE0027H2_tex}.
It can be seen on Fig.~6 of \citet{Lacour05} that their model approximately reproduces the values
measured here. However, their model
% but in their case, the authors stress this process
requires a formation rate $\sim$10 times higher than that measured in
the Galactic ISM \citep{Jura75,Gry02}. It is clear that at the low
metallicity measured in the present DLA system, [Zn/H]~=~$-$1.63, 
this assumption can be rejected.

%Note that the highest rotational levels have a radiative life-time
%considerably shorter than the lowest J levels. This implies that the low
%rotational levels have more time than high-J molecules to cool down by
%collision with \HI, before absorbing a Lyman photon \citep[see,
%  e.g.][]{Spitzer73, Lacour05}. This explains also the non-Boltzman
%distribution of the J~$>2$ levels as seen in Fig.~\ref{HE0027H2_tex}
%and should put an upper limit on the $n_{\rm H}$ density. 
%\citet{Spitzer73} note the lifetime of the J~=~5 level is about
%$\tau \simeq$~10$^8$~s, that is less than the collisional time (at kinetic
%temperature $T\sim100$~K) with H atoms if the proton density is less than about 10~cm$^{-3}$.
%Detailed modeling of the absorbing cloud is under the scope of a following paper.

Another explanation could be that the cloud is composed of several layers
with a gradient of temperature. External layers would be warm 
and exposed to strong external UV flux. In these layers, the excitation of H$_2$
could be large and mostly due to UV pumping. They would contribute
mostly to the column densities of high J-levels.
The internal layers would be cold and shielded from the external UV flux.
It must be realized that the increased $b$ with higher J is coupled with
an excitation diagram described by two excitation temperatures,
$T_{\rm 0-J} \sim 90$~K for J~=~0-2 and $T_{\rm 2-J} \sim 500$~K for J~=~3-6 (see Fig.~\ref{HE0027H2_tex}). This kind of diagram
has been explained in the Galactic ISM by invoking the association of
a diffuse cold cloud with a warm Photo-Dissociation Region (PDR), see \citet{Boisse05}.
%The same model could probably explain the line of sight toward HE\,0027$-$1836.
%A detailed model is however out of the scope of this paper and
%will be the subject of a forthcoming study.

%The dependency of $b$ on $J$ can also be reproduced if we are in case of turbulences, or
%C-shocks. Indeed, the small velocity shift seen between the \CI\ and
%H$_2$ component may also be due to a low-velocity shock, a high
%velocity shock (J-type), in turn, would likely produce a shift between
%the H$_2$ components in different rotational levels, what is not observed. 
%Collisional excitation in a warm environment should
%imply a low H$_2$ formation rate on dust grains and a low molecular fraction. 
%\citet{Lehner03} show that this kind of medium in the local bubble
%has molecular fractions of about $f \sim 10^{-5}$, what is in agreement with
%what is observed for the present system ($f \simeq 7^{+3}_{-2} \times 10^{-5}$).
%Such low molecular fraction are generally observed in our Galaxy on lines of
%sight with low \HI\ column densities ($\log N$(\HI)~$\leq$~20). 
%
%
%%%%%%%%%%%%%%%%%%%%%%%%%
%SECTION FRANCK MODELS
%%%%%%%%%%%%%%%%%%%%%%%%%%
 \subsection{Models}
\newcommand{\CD}{C$_{\rm D}$}
We used the Meudon PDR code \citep{Lepetit06, Goicoechea07} to model 
%component \#3 
the system toward HE\,0027$-$1836 where H$_2$ is detected in
rotational levels J~=~0 to J~=~5. 
This PDR model assumes a stationary plan-parallel slab of dust and gas of constant
hydrogen density $n_{\rm H}$, illuminated by a UV radiation field
and solves the radiative transfer, chemistry and thermal balance.
%
%The level dependent photo-dissociation or photo-ionization of H$_2$ and C~{\sc i} as well 
%as their excitation in 
%ro-vibrational levels and fine structure levels due to radiative excitation, decay and collisional excitation 
%are computed. 
The code used in this paper is a slightly modified version compared to the online 
one\footnote{\url{http://aristote.obspm.fr/MIS}}. %  Since we have no information on the UV spectrum impinging on the 
%cloud 
We adopt the Interstellar Radiation Field (ISRF) as given by \citet{Draine78}, scaled by a factor $\chi$. 
%(A rajouter si ncessaire pour faire le lien avec le reste du papier : 
Note that the energy between 912 and 2400 ${\rm \AA}$ for the Draine ISRF is
1.78 times that of the Habing radiation field. The grain size distribution is assumed to 
follow the \citet{Mathis77} law with radii between 0.1 and 0.3~$\mu$m. The mean 
Galactic dust extinction curve is used. Elemental abundances for C, N, O and S are scaled from solar 
abundances,  [X/H$]_\odot$, of \citet{Morton03}, using the observed sulfur metallicity, [S/H$]$ 
(see Table~\ref{HE0027_metalsT}). 
%Since S is not depleted 
%on dust \citep{Savage96} this gives a crude estimation of the elemental abundances in the DLA. Then 
%
\citet{Savage96} have shown, in the context of diffuse Galactic clouds, that depletion 
varies significantly from one line of sight to another. Here, we test two extreme assumptions to fix 
the gas phase abundances of C and O: the same depletion on dust as in the cool medium towards 
$\zeta$ Oph \citep{Savage96} or no depletion. This gives in the first case 
C/H = $2.35 \times 10^{-6}$ and O/H = $3.9 \times 10^{-6}$ and in the second case 
$6.2 \times 10^{-6}$ and $1.0 \times 10^{-5}$, respectively. 
%This is important only for models where thermal balance is computed.
%Gas phase abundances are obtained assuming the same depletion on dust than towards $\zeta$ Oph \citep{Savage96}
%leading to 
%Since \MgII\ is observed, the observed Mg/H is used. Adopted gas phase elemental abundances are then 
%2.35 10$^{-6}$, 3.9 10$^{-6}$, 3.0 10$^{-7}$, 1.8 10$^{-6}$ for C/H, O/H, S/H and Mg/H respectively. 
CMB temperature is assumed to be 9.2 K ($z = 2.4$).
%Some observed species in \#3 (Mg II, S II) could not be all localized in the same cloud than H$_2$. 
%Moreover, considering the uncertainties in the parameters of the models as the elemental abundances, 
%the main constraints on models are N(H I) which gives a maximum value, the H$_2$ rotational levels 
%and C, C$^+$ abundances and excitations.  \\
%
\par We built two grids of models: one corresponding to isothermal models
with $T_{\rm gas}$ = 130 K corresponding to the observed
value of $T_{\rm 0-1}$ and one corresponding to models solving the thermal balance
equations. For each grid we compute models with
$\chi$~=~ 1, 10, 20, 30 and n$_{\rm H}$~=~ 10, 100, 150, 200, 400, 500 cm$^{-3}$. 
We checked the influence of the dust-to-gas mass ratio, $g$, or equivalently 
\CD\ = $N_{\rm H}$/E(B-V), the ratio of total neutral hydrogen
column density to color excess, using $g$ = 0.01 and \CD\ = $5.8 \times
10^{21}$ \citep[corresponding to the ISM values;][]{Bohlin78}, 0.001, $5.8 \times 10^{22}$ 
and  0.0001, $5.8 \times 10^{23}$~cm$^{-2}$\,mag$^{-1}$. 
Column densities are computed 
perpendicular to the plane parallel slab. Since the size of the cloud is not known, all models 
assume a semi-infinite geometry up to reproduce half of the observed $N$(H$_2$). Then column 
densities are multiplied by 2 to simulate a slab of gas illuminated on both sides. The model is 
constrained by the condition that the total observed $N$(H$_2$)
should be reproduced.
\par
\begin{figure}[t]
\includegraphics[angle=-90,width=\hsize]{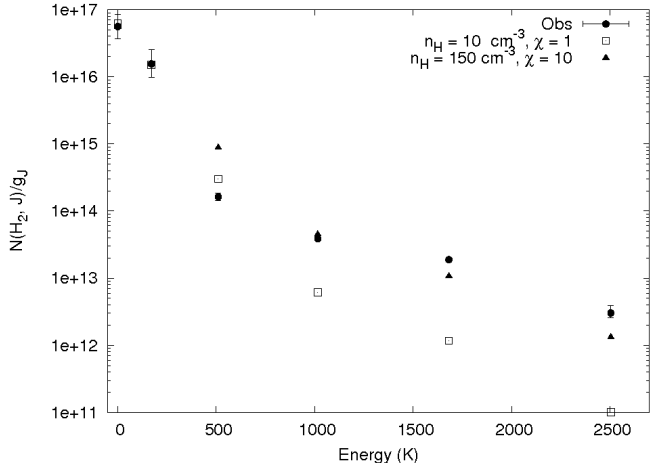}
\caption{Excitation diagrams obtained for isothermal models with
  $T_{\rm gas}$~=~130\,K. 
Grain parameters are $g$ = 0.001 and \CD\ = $5.8 \times 10^{22}$~cm$^{-2}$\,mag$^{-1}$.}
\label{fig:diagexT130}
\end{figure}

 The best iso-thermal models are obtained with $g$ = 0.001 and 
\CD\ = $5.8 \times 10^{22}$~cm$^{-2}$\,mag$^{-1}$
corresponding, as expected, to a low dust content (note that if we scale depletion
with the observed [Zn/Cr] ratio, see previous Section, we obtain $g$~=0.002). 
Figure \ref{fig:diagexT130} presents the 
excitation diagrams obtained with isothermal assumption in two cases. 
The best fit to the data for all levels is obtained with
$n_{\rm H}$~=~150~cm$^{-3}$, $\chi$~=~10 (about 20 times the
Habing radiation field). This model reproduces reasonably well all column densities
except that of the J~=~2 level which is over-estimated by nearly an order
of magnitude. The low observed value of $N$(J=2) relative to other column densities
is the special feature of this DLA system.
%the excitation of all observed H$_2$ levels except J = 2 which is over-estimated
%by a factor 5. 
In this model, the molecular fraction is $f = 10^{-4}$
so H is the main collisional partner for H$_2$. It must be noted that H/H$_2$ collision
rates are poorly known mainly because of reactive collisions. However
using collision rates from \citet{Flower98b} instead of those
from \citet{Mandy93} does not change the results significantly.
Amongst models of total H$_2$ column density corresponding to the
observed one, only those with low-density, $n_{\rm H}$ $\simeq$~10\,cm$^{-3}$,
and weak radiation field, $\chi \simeq 1$ (Fig. \ref{fig:diagexT130}),
can reproduce the observed $N$(H$_2$, J=2).
% can be reproduced only by low-density models, $n_{\rm H}$ $\simeq$~10\,cm$^{-3}$,
%and weak radiation field, $\chi \simeq 1$ (Fig. \ref{fig:diagexT130}). 
Higher densities tend to thermalize J~=~2
whereas higher radiation fields tend to pump this level too much.
%With the above parameters however populations of J~$>$~2 levels are
%under-estimated by more than one order of magnitude. 
The influence of the distribution of H$_2$  amongst ro-vibrational levels after 
formation on dust has also been tested. 
%Assuming ro-vibrational excitation corresponding 
%to a Boltzman distribution at $T_{\rm gas}$, a H$_2$ formation in
%vibrational level $\nu$~=~6 or in rotational level J~=~14 
%does not have any significant influence on the final $N$(H$_2$, J=2). 
%So we can confidently conclude that the low observed $N$(H$_2$, J=2) is a probe of a 
%low-density medium illuminated by a weak radiation field. 
It can be noted that a few 
lines of sight with $T_{0-2}$ $\le$ $T_{0-1}$ have been observed by FUSE and Copernicus 
in our Galaxy or the Magellanic clouds. The HD~108927 line-of-sight has been modelled 
by \citet{Gry02}. The authors reached similar conclusions.
%also concluded on a low density medium (28 cm$^{-3}$) and a low radiation field $\chi$ = 1. 
However, with such low density and UV flux ($n_{\rm H} \simeq 10$~cm$^{-3}$, $\chi=1$), 
populations of $J > 2$ levels will be under-estimated by more than one order of magnitude.
Note, however, that in these models, $N$(\CI) and $N$(\CI$^*$) are in good agreement with 
observations. %(Tab. \ref{tab:modgrains}, model M2). 
Depending on the adopted depletion of C into dust grains, 
$N$(\CII$^*$), $N$(\CI) and $N$(\CI$^*$) are reproduced 
within a factor of 2 to 3 which is acceptable considering the uncertainties in the model.
% (Table \ref{tab:modgrains}). 
$N$(\SII), $N$(\MgI) and $N$(\MgII) are also in good agreement with observations. 
\begin{figure}[!h]
\includegraphics[clip=,angle=-90,width=\hsize]{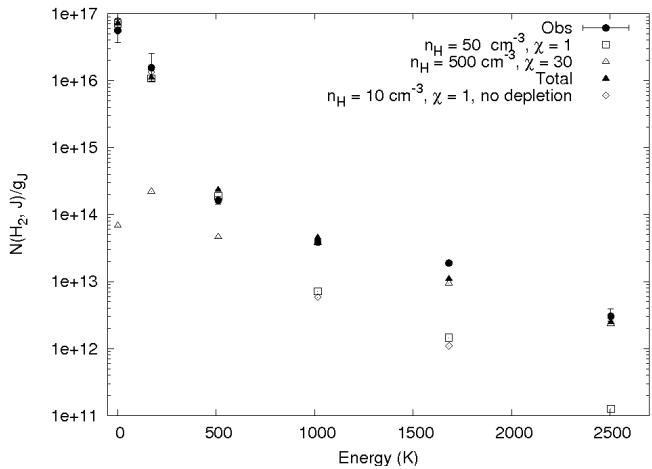}
\caption{Excitation diagrams obtained with thermal balance models. Grains parameters are 
$g$ = 0.001 and \CD\ = $5.8 \times 10^{22}$~cm$^{-2}$\,mag$^{-1}$.}
\label{fig:diagexbilth}
\end{figure} 
\par Thermal balance models lead to similar conclusions. These models are, in principle, more
constrained as they have to reproduce $T_{0-1}$. However, they are dependent on assumptions on (i) the dust composition since the photo-electric effect on dust is the main heating mechanism and (ii) the elemental abundance of oxygen which controls cooling. 
Best model is again for $g$ = 0.001 and \CD\ = $5.8 \times 10^{22}$ cm$^{-2}$\,mag$^{-1}$.
As for isothermal models, a low density is required to avoid to overestimate $N$(H$_2$, J=2). 
With the first set of abundances (same depletion as in the cool medium towards $\zeta$ Oph), 
the best model is $n_{\rm H}$ = 50 cm$^{-3}$ and $\chi$~=~1. In this case, $T_{0-1}$ = 90 K. 
A lower density of 10 cm$^{-3}$ increases $T_{0-1}$ to 180 K because of
the low abundance of coolants. 
With the second set of abundances (no depletion), $n_{\rm H}$ = 10 cm$^{-3}$ and $\chi$ = 1 is favored 
and gives $T_{0-1}$~=~96~K. The corresponding excitation diagrams are presented in 
Fig. \ref{fig:diagexbilth}.
\par In the two models, $T_{0-1}$ is representative of the kinetic temperature which varies from 
about 100 to 88 K when the depth in the cloud increases. The best-fit model
with a depletion following that towards $\zeta$ Oph 
gives $N$(\CII$^*$) = $5 \times 10^{13}$, $N$(\CI) = $8.0 \times 10^{11}$ and 
$N$(\CI$^*$) = $7.8 \times 10^{11}$ cm$^{-2}$. The one with no
depletion gives $1.4 \times 10^{14}$, $5.8 \times 10^{12}$ and $3.2
\times 10^{12}$ cm$^{-2}$ for the same species. Both
models are within a factor of two from observations (see Tables \ref{HE0027_CIIstart} and \ref{HE0027H2CI}).
%These two sets of gas phase abundances surround in a factor 2 the observations. 
%
%
\par The population of high-J levels cannot be reproduced by the above low-density 
models and a PDR component has
to be added. High density and strong radiation field are required in order
to explain H$_2$ excitation by fluorescence.
%To reproduce column densities of H$_2$ in J $>$ 2 by a PDR component, high density and high 
%radiation field are required. 
Fig. \ref{fig:diagexbilth} gives an example in which the high-J populations are 
produced in a clump of density $n_{\rm H}$ = 500 cm$^{-3}$ 
with $\chi = 30$. The corresponding size is 0.3 pc. 
Note that the value of the UV radiation field intensity
is consistent with what is estimated analytically in Sects.~\ref{secH2ex} and \ref{secCII}.
\par However, the probability that the line of sight crosses such a clump is small especially when 
the model requires that this small clump must be embedded in a particularly strong
UV radiation field. Note also that there is no velocity shift between the absorption
lines of the different J levels suggesting that they are not produced in very
different locations.
%It would be more reasonable to believe the H$_2$ excitation 
Another possibility to explain the excitation is turbulent dissipation
either as vortices or C-shocks \citep{Joulain98, Cecchi-Pestellini05, Gredel02, LePetit04}. 
In that case, increased temperature in turbulent vortex 
would be responsible for the excitation of the higher H$_2$ rotational levels.
Indeed, \citet{Joulain98} show that there should be no shift between the centroids
of different species and that the widths of the lines should be larger for higher
excitation as observed in the present case.
\citet{Cecchi-Pestellini05} show that a small amount of hot gas 
located in turbulent dissipative cells can explain the H$_2$ excitation.
However, temperature excitation as high as $T_{\rm 2-J}$~=~500~K for a low 
total $N$(H$_2$, J$>$2) may be difficult to be reproduced by such models.
This assumption probably deserves more detailed investigation.

%can only be reproduced by excitation by a high 
%UV radiation field.
%However temperatures usually quoted are below 300K. This is less
%than our temperature (see Falgarone et al. 2005, A\&A, 433, 997 and Gry et al.
%A\&A, 2002, 391, 675)
 
% %%%%%%%%%%%%%%%%%%%%%%%%%%%%%%%%%%%%%%%%%%%%%%%
% END MODELS
%%%%%%%%%%%%%%%%%%%%%%%%%%%%%%%%%%%%%%%%%%%%%%%%

\section{Comments on metallicity and depletion}

\begin{figure}[!Ht]
 \begin{center}
 \includegraphics[clip=,width=0.98\hsize]{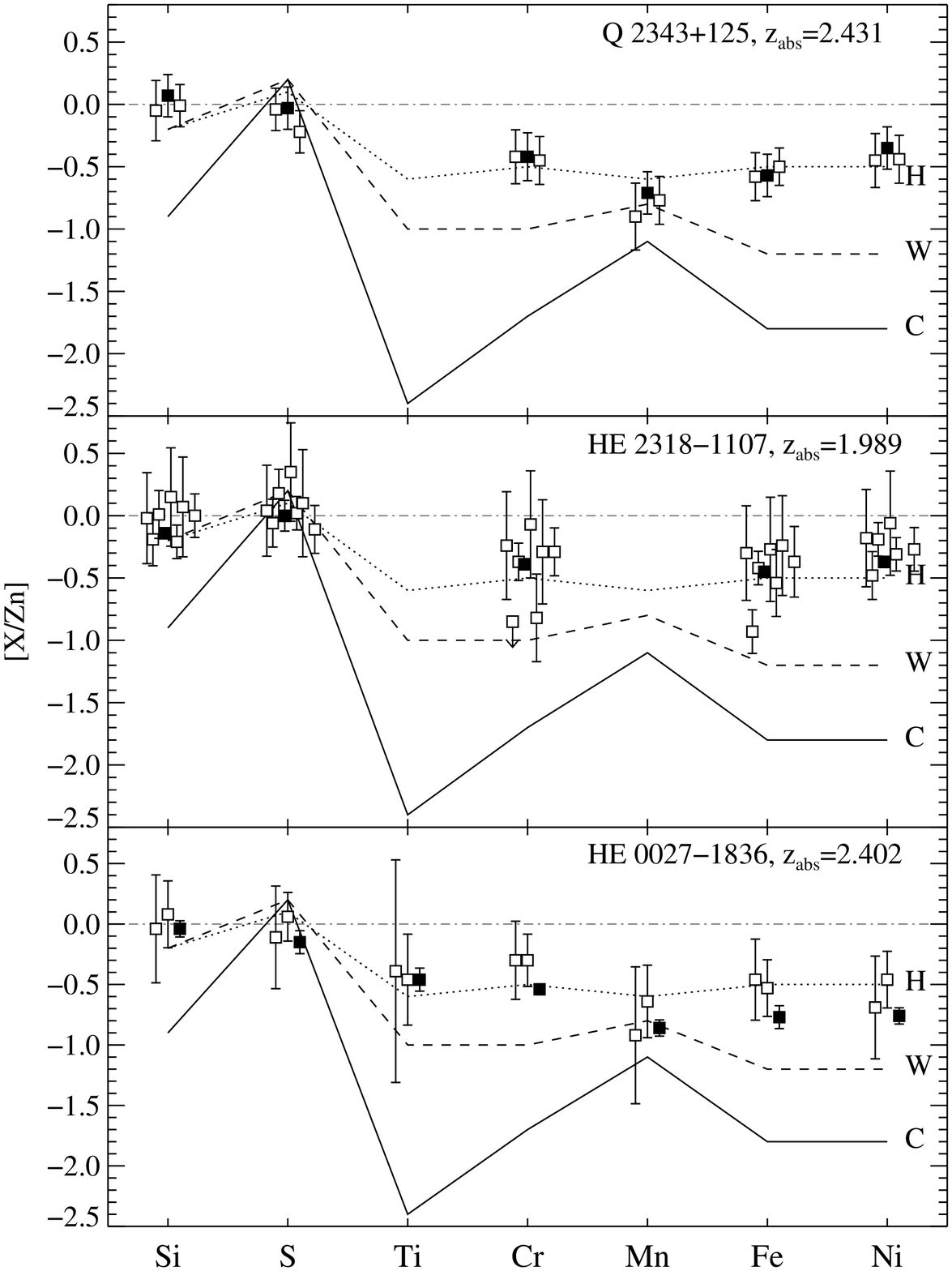}
 \caption{\label{pattern} Depletion pattern for the three systems
 presented in this paper. The solid, dashed and dotted lines,
 represent the typical relative abundances observed in, respectively, 
cold (C) and warm (W) gas in the Galactic disk and diffuse gas
in the Galactic halo (H), from \citet{Welty99}. The dashed-dotted line corresponds
 to solar abundances.
 Filled squares stand for H$_2$-bearing components, open squares for other
components in the system. We show only the components in which \ZnII\
 is detected. For each system, components are shifted 
in abscissa according to their redshift (i.e., the bluest components are on the left). 
It is apparent that the Galactic halo depletion pattern represents best the observed
 abundances. There is no significant difference
 between H$_2$-components and the other components although it seems that
 depletion (from \CrII, \MnII, \FeII, and \NiII) is slightly higher
 in the H$_2$ component of the DLA toward HE\,0027$-$1836.}
 \end{center}
\end{figure}

\begin{figure}[!h]
 \begin{center}
 \includegraphics[clip=,width=0.95\hsize]{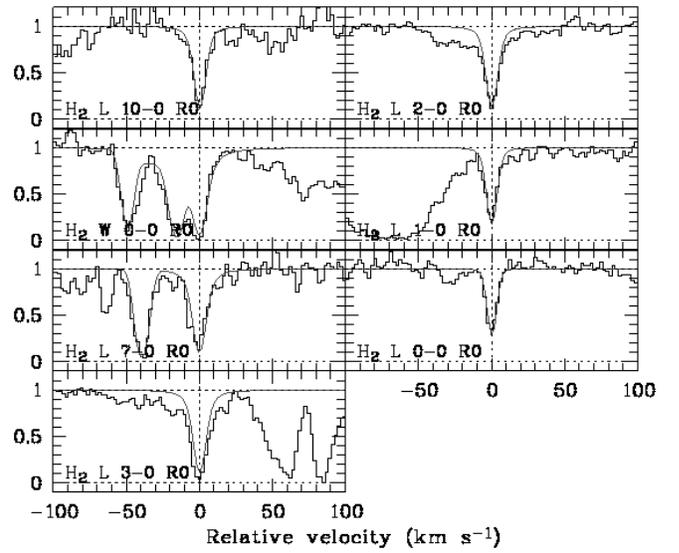}
 \caption{Absorption lines from H$_2$ in rotational level J~=~0 at
 $\zabs=2.40183$ toward HE\,0027$-$1836. 
%{\bf The origin (v~=~0~km\,s$^{-1}$) of the velocity scale
% corresponds to $z=2.40186$, where the metals and CI absorptions are
% the strongest.}
\label{HE0027_J0}}
 \end{center}
\end{figure}

\begin{figure}[!h]
 \begin{center}
 \includegraphics[clip=,width=0.95\hsize]{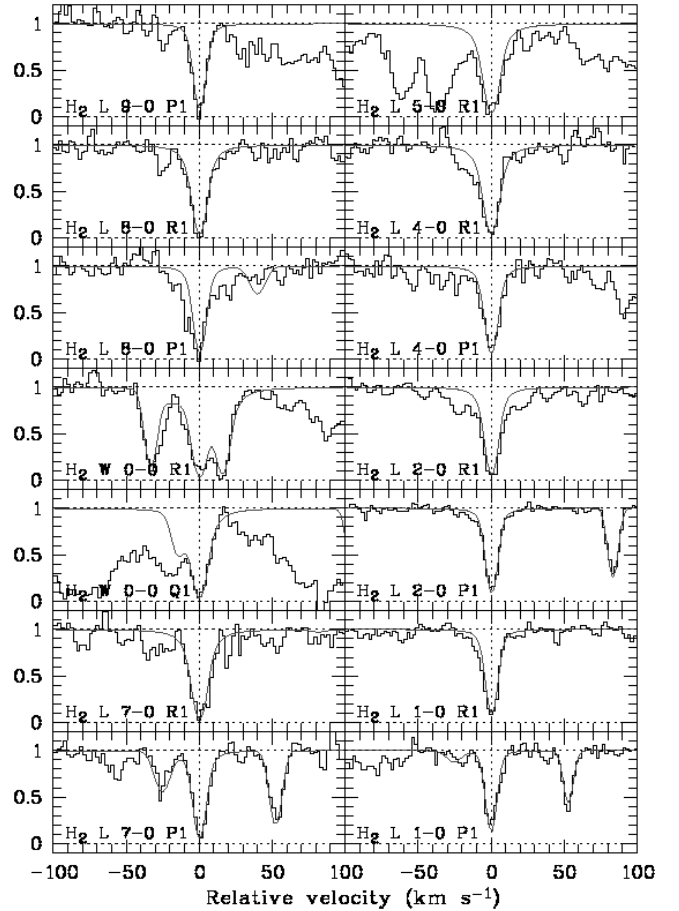}
 \caption{Absorption lines from H$_2$ in rotational level J~=~1 at $\zabs=2.40183$ toward
 HE\,0027$-$1836. One can see that damping wings are present for 
H$_2$\,L8-0\,R1. \label{HE0027_J1}}
 \end{center}
\end{figure}

\begin{figure}[!h]
 \begin{center}
 \includegraphics[clip=,width=0.95\hsize]{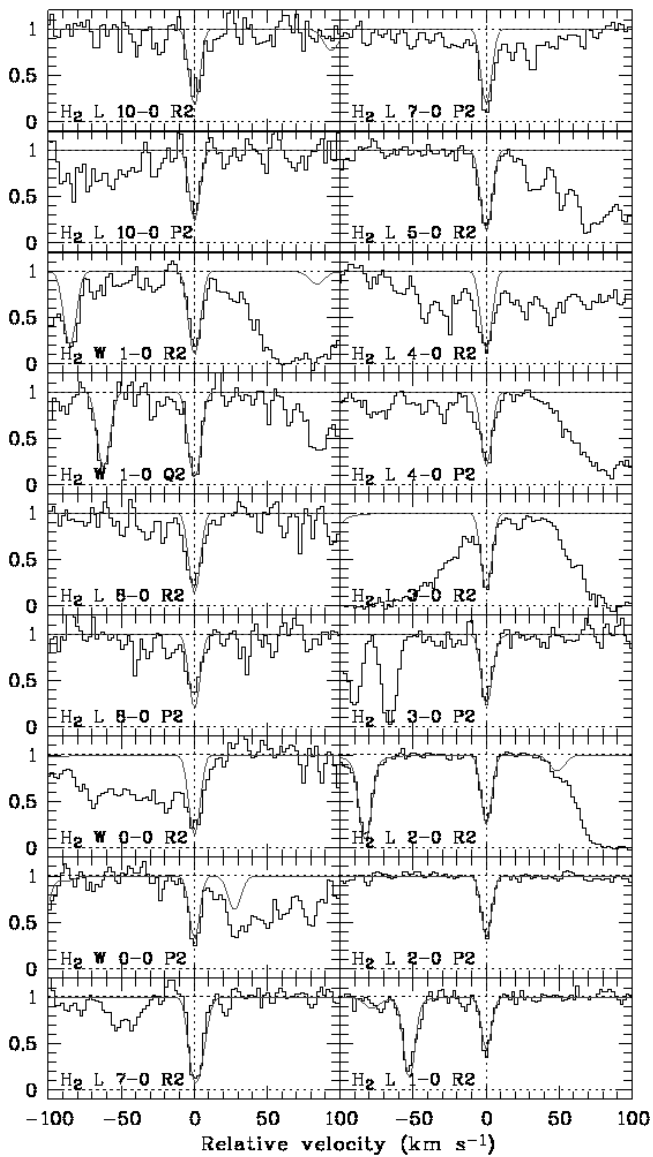}
 \caption{Absorption lines from H$_2$ in rotational level J~=~2 at $\zabs=2.40183$ toward
 HE\,0027$-$1836. Note the high signal-to-noise ratio in the
 H$_2$\,L2-0\,R2 and  H$_2$\,L2-0\,P2 panels.  \label{HE0027_J2}}
 \end{center}
\end{figure}

\begin{figure}[!h]
 \begin{center}
 \includegraphics[clip=,width=0.95\hsize]{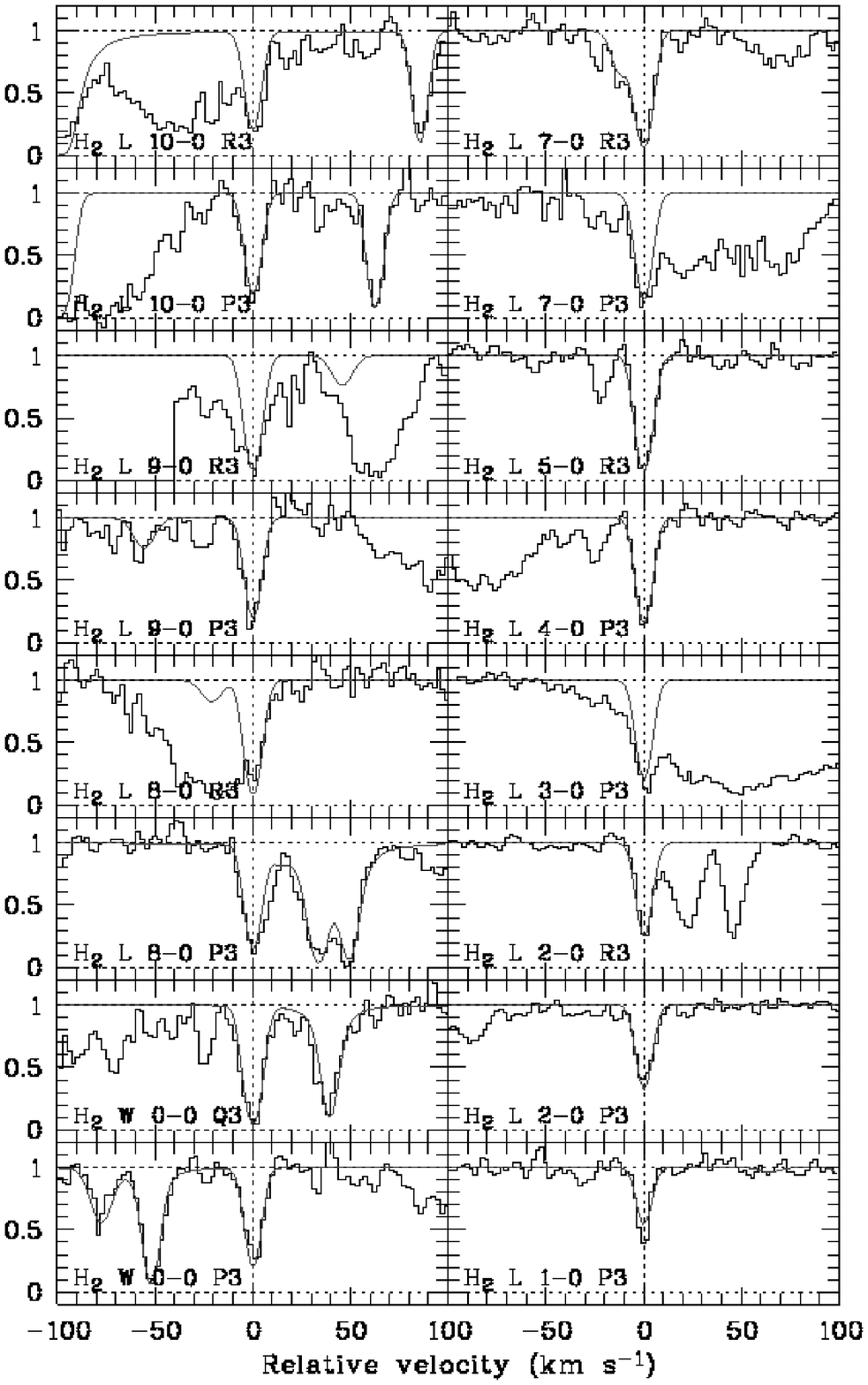}
 \caption{Absorption lines from H$_2$ in rotational level J~=~3 at $\zabs=2.40183$ toward
 HE\,0027$-$1836. \label{HE0027_J3}}
 \end{center}
\end{figure}

\begin{figure}[!h]
 \begin{center}
 \includegraphics[clip=,width=0.95\hsize]{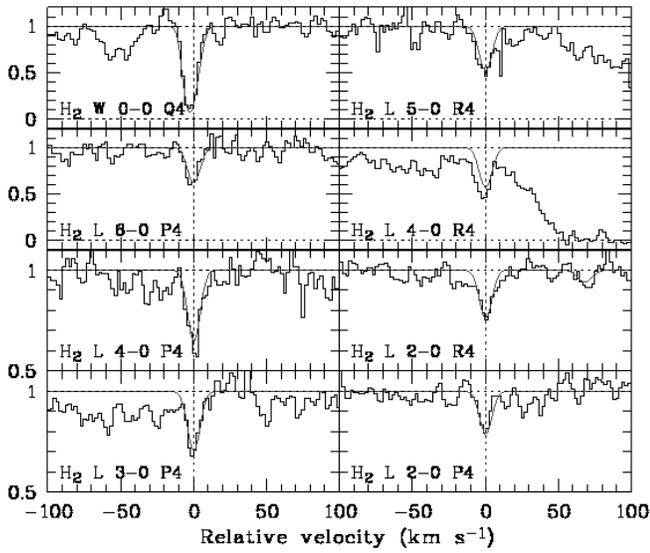}
 \caption{Absorption lines from H$_2$ in rotational level J~=~4 at $\zabs=2.40183$ toward
 HE\,0027$-$1836. The  H$_2$\,W0-0\,Q4 transition  is actually blended with H$_2$\,L7-0\,R2 which provides most of the absorption. \label{HE0027_J4}}
 \end{center}
\end{figure}

\begin{figure}[!h]
 \begin{center}
 \includegraphics[clip=,width=0.95\hsize]{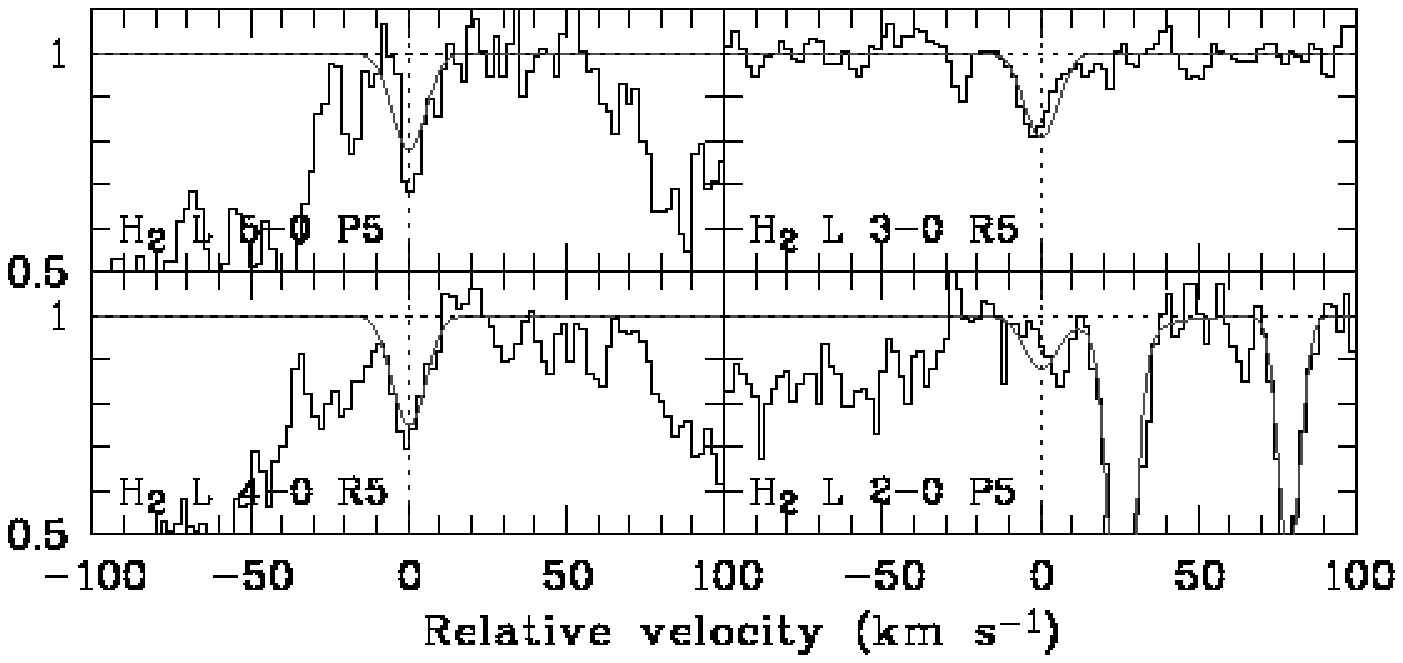}
 \caption{Absorption lines from H$_2$ in rotational level J~=~5 at $\zabs=2.40183$ toward
 HE\,0027$-$1836.\label{HE0027_J5} }
 \end{center}
\end{figure}

\begin{figure}[!h]
 \begin{center}
 \includegraphics[clip=,width=0.95\hsize]{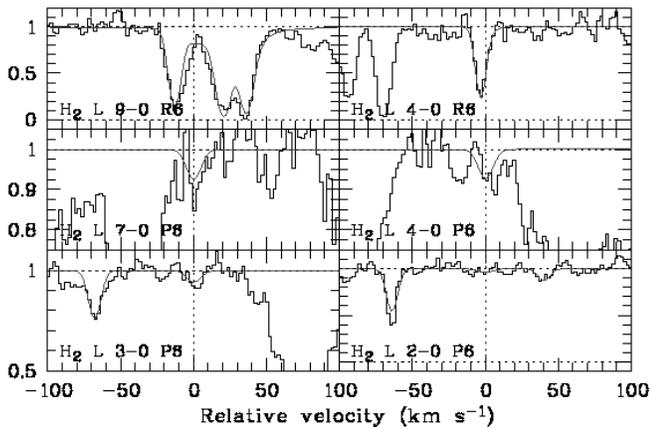}
 \caption{Possible detection of H$_2$ in rotational level J~=~6 at $\zabs=2.40183$ toward
 HE\,0027$-$1836. A consistent absorption feature is seen in the
 H$_2$\,L3-0\,P6 panel, and explains the slight asymmetry in the
 H$_2$\,L4-0\,R6 panel, which actually is a blend of
 H$_2$\,L3-0\,P2 together with H$_2$\,L4-0\,R6.  \label{HE0027_J6} }
 \end{center}
\end{figure}

 In Fig.~\ref{pattern}, we compare the depletion patterns for the three
 systems presented here to the typical depletions observed in 
 cold and warm gas of the Galactic disk and gas in the Galactic halo.
The different points correspond to the different components in each
 system in which \ZnII\ is detected.
It is apparent that the Galactic halo depletion pattern represents best the observed
 abundances.
% The depletion in the H$_2$ component seems to be consistently slightly higher for the
% system toward HE\,0027$-$1836, but the difference with the other components is quite small.
The differences between components are small.
In the DLA system toward HE\,0027$-$1836, the depletion found in the H$_2$-component 
is the highest in the system however: [Zn/Cr$]$~=~0.54, while it is $\sim$~0.3 in 
the other components. Similarly, [Zn/Fe$]$~=~0.77 in the H$_2$-bearing component, while
[Zn/Fe$]=$~0.46 and 0.53 in the other two components.        
Note that silicon is almost non-depleted in all components of
 the three systems. This is similar to what is observed through some lines of
 sight in the Small Magellanic Cloud \citep{Welty01}. 
%It is therefore not a good probe of the depletion factors.%
 The small depletion factor, together with low metallicities, 
%it is clear 
implies that the dust content is
 small and can explain the low observed molecular fraction. The fact that we do 
not detect H$_2$ in other components 
 cannot only be due to the little difference in dust-to-gas ratio,
 resulting in a lower H$_2$ formation rate. The main reason is probably the lower
 column densities in the components (especially for HE\,2318$-$1107).
%, and therefore lower dust- and mostly self-shieldings. 
%
We summarize the total abundances found in
the three systems in Table~\ref{sum_table}. 
%Note that we have integrated the metal column density over the whole
%profiles (see, however, \S\,\ref{parq2343m}).
%Metallicities might be less in case we restrict this integration to
%the components where H$_2$ and \CI\ are detected. 

%In the case of
%HE\,2318$-$1107 this would decrease metallicities by a factor of 3 to 4 to 
%
% a central and strongest component in which H$_2$ and \CI\ are detected. 
%Considering 
%that this components probably contains most of the neutral hydrogen, the abundances would be 
%decreased by a factor 3 to 4, then 
%[Si/H$]=-1.52$, [S/H$]=-1.34$, [Cr/H$]=-1.77$, [Fe/H$]=-1.83$, 
%[Ni/H$]=-1.75$ and [Zn/H$]=-1.38\pm0.05$.
%In addition, the low expected
% column densities of H$_2$, even if following the metals column densities can also 
%make H$_2$ below the limit of detectability for the systems
% toward HE\,2318$-$1836 and Q\,2343$+$125.
%
% This shows
%that dust still plays a role in governing the presence of H$_2$ in
%this system. Note that
%silicon, in turn, is almost non-depleted in the three components, what can be interpreted as
%few silicates in the dust, similar to what have been observed in some
%lines of sight toward the Small
%Magellanic Cloud \citep{Welty01}.
%
%
\begin{center}
\begin{table*}[!ht]
\caption{\label{sum_table} Summary of metals abundances in newly discovered H$_2$-bearing DLAs.}
\begin{center}
\begin{tabular}{c c c c c c c}
\hline
\hline
QSO                  & \multicolumn{2}{c}{Q\,2343$+$125, $z_{\rm em}=2.51$}   & \multicolumn{2}{c}{HE\,2318$-$1107, $z_{\rm em}=2.96$}  & \multicolumn{2}{c}{HE\,0027$-$1836, $z_{\rm em}=2.56$}  \\
$\zabs$              & \multicolumn{2}{c}{2.431}           & \multicolumn{2}{c}{1.989}            & \multicolumn{2}{c}{2.402}            \\
$\log N(\ion{H}{i})$ & \multicolumn{2}{c}{20.40~$\pm$~0.07}  & \multicolumn{2}{c}{20.68~$\pm$~0.05}   & \multicolumn{2}{c}{21.75~$\pm$~0.10}   \\
$\log N$(H$_2$)      & \multicolumn{2}{c}{13.69~$\pm$~0.09}  & \multicolumn{2}{c}{15.49~$\pm$~0.03}   & \multicolumn{2}{c}{17.30~$\pm$~0.07}   \\  
$\log f$             & \multicolumn{2}{c}{$-$6.41~$\pm$~0.16}& \multicolumn{2}{c}{$-$4.89~$\pm$~0.08} & \multicolumn{2}{c}{$-$4.15~$\pm$~0.17} \\
\hline
Ion (X)              & $\log N$(X)        &   [X/H$]$          & $\log N$(X)      &   [X/H$]$         &  $\log N$(X)      &   [X/H$]$        \\
\hline
%\CI                 & $<$~12.1         &                  &                  &                   &12.25$^{+0.09}_{-0.15}$&                  \\
%\CI$^*$             &                  &                  &                  &                   & $\leq$12.27       &                  \\
\NI                  & 14.62~$\pm$~0.01   & $-$1.73~$\pm$~0.07 & $>$~14.55$^a$    & $>-$2.08$^a$     & 15.25~$\pm$~0.08    & $-$2.45~$\pm$~0.13 \\
\MgI                 & 12.40~$\pm$~0.15   &                    & 12.73~$\pm$~0.11    &                     & 12.58~$\pm$~0.04    &                  \\
\MgII                & $<$~14.69          & $<-$1.29           &$>$~14.94$^b$ & $>-$1.32$^b$& 16.00~$\pm$~0.04  & $-$1.33~$\pm$~0.11 \\
\SiII                & 15.15~$\pm$~0.03   & $-$0.81~$\pm$~0.08 & 15.34~$\pm$~0.01    & $-$0.90~$\pm$~0.05  & 15.67~$\pm$~0.03    & $-$1.64~$\pm$~0.10 \\
\PII                 & 13.05~$\pm$~0.03   & $-$0.91~$\pm$~0.08 &    ---              &    ---              & 13.17~$\pm$~0.23    & $-$2.14~$\pm$~0.25 \\
\SII                 & 14.66~$\pm$~0.02   & $-$0.94~$\pm$~0.07 & 15.09~$\pm$~0.02    & $-$0.79~$\pm$~0.05  & 15.23~$\pm$~0.02    & $-$1.72~$\pm$~0.10 \\
\ArI                 & 13.19~$\pm$~0.01   & $-$1.61~$\pm$~0.07 &    ---              &    ---              & 14.42~$\pm$~0.02    & $-$1.73~$\pm$~0.10 \\
\TiII                & $<$~11.85          & $<-$1.49         & $<$~12.00             & $<-$1.62            & 12.61~$\pm$~0.06    & $-$2.08~$\pm$~0.12 \\
\CrII                & 12.87~$\pm$~0.03   & $-$1.22~$\pm$~0.08 & 13.13~$\pm$~0.06    & $-$1.24~$\pm$~0.08  & 13.37~$\pm$~0.01    & $-$2.07~$\pm$~0.10 \\
\MnII                & 12.35~$\pm$~0.03   & $-$1.58~$\pm$~0.08 &    ---           &     ---                & 12.84~$\pm$~0.03    & $-$2.44~$\pm$~0.10 \\
\FeII                & 14.52~$\pm$~0.02   & $-$1.38~$\pm$~0.07 & 14.91~$\pm$~0.01    & $-$1.27~$\pm$~0.05  & 14.97~$\pm$~0.02    & $-$2.28~$\pm$~0.10 \\
\NiII                & 13.43~$\pm$~0.03   & $-$1.22~$\pm$~0.08 & 13.82~$\pm$~0.03    & $-$1.11~$\pm$~0.06  & 13.70~$\pm$~0.02    & $-$2.30~$\pm$~0.10 \\
\ZnII                & 12.20~$\pm$~0.07   & $-$0.87~$\pm$~0.10 & 12.50~$\pm$~0.03    & $-$0.85~$\pm$~0.06  & 12.79~$\pm$~0.02    & $-$1.63~$\pm$~0.10 \\

\hline

\end{tabular}
\end{center}
\footnotesize
$^a$ Because of many blends, we take $N$(\NI) in the central component as a lower limit
on the total neutral nitrogen column density.\\
$^b$ Considering only $N$(\MgII) in the main component since the other
components are either blended with \lya\ forest absorptions or not detected.
%, their
%contribution to $N$(\MgII) should be negligible, with similar
%upper-limits to that on components 6, 7, and 8 (see Table
%\ref{HE2318_metalsT}).
\normalsize
\end{table*}
\end{center}
%PUT Q0347??
%\begin{figure}
% \begin{center}
% \includegraphics[clip=,width=\hsize]{pattern347.eps}
% \caption{\label{pattern347}}
% \end{center}
%\end{figure}
\section{Conclusion}
We have presented a detailed analysis of three H$_2$-bearing damped Lyman-$\alpha$
systems. Two of them are reported here for the first time. This brings the number of known 
high-redshift ($\zabs>1.8$) H$_2$-bearing DLAs to twelve.
%, considering the one at $\zabs=3.390$ toward Q\,0000$-$2620
%\citep{Levshakov00} as a firm detection. 
All three systems have low-molecular fractions ($\log f \leq
-4$). Only one DLA system with such a low 
molecular fraction was reported before, at $\zabs=3.025$ toward 
Q\,0347$-$383 \citep{Levshakov02}.
\par The depletion patterns observed in the components of the three systems are 
very similar to what is observed in gas located in the Galactic halo,
probably because of similarly low metallicities. The depletion is 
not very different, although slightly larger, in the H$_2$-bearing components
compared to the other components in the systems.
This is different from what is seen in systems with larger molecular fractions
where large depletion factors are usually observed in H$_2$-bearing
components \citep{Ledoux03, Rodriguez06}. This could be a
consequence of the relation between molecular fraction and metallicity \citep{Petitjean06}.
\par The system toward HE\,0027$-$1836 is particularly interesting as it
shows absorption from rotational levels J~=~0 to 5 (and possibly J~=~6)
in a single well-defined component.
The UV radiation field intensity in which this system is immersed is found to
be about 20 times the Galactic ambient flux. This could be an upper limit as 
it is estimated  assuming that H$_2$ excitation is mainly due to UV pumping.
However the same value is found when estimating the cooling rate 
in the gas from the \CII$^*$ column density. 
\par Thanks to the very high data quality, we observe
for the first time at high $z$ an increase of the Doppler parameter $b$ 
from low to high H$_2$ rotational levels. %Doppler parameters are higher for higher J. 
%As a first result, we suggest observers should pay attention to a possible
%dependency between $b$ and $J$, especially when fitting saturated lines.
%The relation $b=f(J)$ can be easily fitted if considering the Doppler
%broadening is only due to different kinetic energies for each
%rotational level. 
To the first approximation, there is a linear relation between the kinetic energy
of the molecule (as given by the Doppler parameter) and the energy of the rotational levels. 
The explanation that this is a direct consequence
of the formation of H$_2$ onto dust-grains \citep[see][]{Lacour05} is
difficult to accommodate given the low formation rate in the DLA gas
due to low metallicity and small dust-to-gas ratio.
% An equilibrium between formation on dust
%grains, UV pumping and photo-dissociation would therefore be at play
%and should strongly depend on the particle density. 
%
\par Detailed PDR models have been constructed to reproduce the
observations in this system that is special because of the low $N$(H$_2$,J=2) 
column density compared to other J level
column densities. Two components are needed, one with low particle density and
weak radiation field to reproduce J~$\leq$~2 levels column densities and one with 
high density and strong radiation field for J~$>$~2 levels. 
This combination may appear ad-hoc but is unavoidable in the context
of PDR models.
On the other hand, the Doppler broadening and the excitation of H$_2$
can also be explained by increased temperature in part of the gas
due to turbulent dissipation or
C-shocks \citep{Joulain98,Cecchi-Pestellini05,Lacour05}, as supported
by the small depletion, the low molecular fraction and the small, if any, velocity shift 
between H$_2$ and \CI\ absorptions.
%In both explanations however, the particle density cannot be very
%high, and is likely to be below 10~cm$^{-3}$, although this has to be confirmed.
%The origin of the H$_2$ lines broadening is therefore still unclear, and
%requires detailed modeling, to choose between one of the two proposed
%mechanisms.
%Detailed models should be constructed to explain this situation probably
%associating diffuse cold gas with a PDR to explain the presence of
%gas with different temperatures and different UV excitations. 

\acknowledgement{We thank the anonymous referee for useful comments. 
PN is supported by a PhD fellowship from ESO. PPJ and
  RS gratefully acknowledge support from the Indo-French Centre for
  the Promotion of Advanced Research (Centre Franco-Indien pour la Promotion de la Recherche Avanc\'ee) under contract No. 3004-3.}

\bibliographystyle{aa}
\bibliography{8021bib.bib}

\end{document}